\documentclass[11pt,a4paper]{article}
\pdfoutput=1
\usepackage{jheppub}
\usepackage{amsmath,amsfonts,amssymb,bbm, mathtools, mathrsfs}
\usepackage{hyperref, csquotes}
\usepackage{graphicx, color, svg,subcaption}
\usepackage{tikz,pgf,tikz-cd,float}
\usepackage{physics,tensor,upgreek}
\usepackage{rotating}
\usepackage[nottoc,numbib]{tocbibind}
\usepackage{accents}

\definecolor{timelike}{RGB}{227, 11, 91}
\definecolor{spacelike}{RGB}{0, 128, 128}
\definecolor{lightlike}{RGB}{0, 25, 150}

\newcommand{\qmarks}[1]{``#1''}
\newcommand*\diff{\mathop{}\!\mathrm{d}}

\begin{document}
\title{Quantum Gravity, Hydrodynamics and Emergent Cosmology: A Collection of Perspectives}

\author[a,b,c]{Jibril Ben Achour,}
\emailAdd{jibril.ben-achour@ens-lyon.fr}

\author[d]{Dario Benedetti,}
\emailAdd{dario.benedetti@polytechnique.edu}

\author[e]{Martin Bojowald,}
\emailAdd{bojowald@psu.edu}

\author[f]{Robert Brandenberger,}
\emailAdd{rhb@physics.mcgill.ca}

\author[g]{Salvatore Butera,}
\emailAdd{salvatore.butera@glasgow.ac.uk}

\author[h]{Renata Ferrero,}
\emailAdd{renata.ferrero@fau.de}

\author[i]{Flaminia Giacomini,}
\emailAdd{fgiacomini@phys.ethz.ch}

\author[h]{Kristina Giesel,}
\emailAdd{kristina.giesel@gravity.fau.de}

\author[a,b]{Christophe Goeller,}
\emailAdd{goellerchristophe@gmail.com}

\author[j]{Tobias Haas,}
\emailAdd{tobias.haas@ulb.be}

\author[k]{Philipp A. Höhn,}
\emailAdd{philipp.hoehn@oist.jp}

\author[l]{Joshua Kirklin,}
\emailAdd{jkirklin@perimeterinstitute.ca}

\author[m,k]{Luca Marchetti,}
\emailAdd{luca.marchetti@oist.jp}

\author[a,n,o]{Daniele Oriti,}
\emailAdd{doriti@ucm.es}

\author[p]{Roberto Percacci,}
\emailAdd{percacci@sissa.it}

\author[q]{Antonio D. Pereira,}
\emailAdd{adpjunior@id.uff.br}

\author[r]{Andreas G. A. Pithis,}
\emailAdd{andreas.pithis@uni-jena.de}

\author[s]{Mairi Sakellariadou,}
\emailAdd{mairi.sakellariadou@kcl.ac.uk}

\author[r]{Sebastian Steinhaus}
\emailAdd{sebastian.steinhaus@uni-jena.de}

\author[t]{and Johannes Thürigen}
\emailAdd{johannes.thuerigen@uni-muenster.de}

\affiliation[a]{Munich Center for Quantum Science and Technology (MCQST), Schellingstr. 4, 80799 M\"unchen, Germany, EU}
\affiliation[b]{Arnold Sommerfeld Center for Theoretical Physics, Ludwig-Maximilians-Universit\"at München, Theresienstrasse 37, 80333 M\"unchen, Germany, EU}
\affiliation[c]{Universit\'{e} de Lyon, ENS de Lyon, Laboratoire de Physique, CNRS UMR 5672, Lyon 69007, France}
\affiliation[d]{CPHT, CNRS, \'{E}cole polytechnique, Institut Polytechnique de Paris, 91120 Palaiseau, France}
\affiliation[e]{Institute for Gravitation and the Cosmos,
The Pennsylvania State University, 104 Davey Lab, University Park, PA 16802, USA}
\affiliation[f]{Department of Physics, McGill University, Montr\'{e}al, QC, H3A 2T8, Canada}
\affiliation[g]{School of Physics and Astronomy, University of Glasgow, Glasgow G12 8QQ, UK}
\affiliation[h]{Institute for Quantum Gravity, FAU Erlangen – Nürnberg,
Staudtstr. 7, 91058 Erlangen, Germany}
\affiliation[i]{Institute for Theoretical Physics, ETH Zürich, 8093 Zürich, Switzerland}
\affiliation[j]{Centre for Quantum Information and Communication, \'{E}cole polytechnique de Bruxelles, CP 165, Universit\'{e} libre de Bruxelles, 1050 Brussels, Belgium}
\affiliation[k]{Okinawa Institute of Science and Technology Graduate University, Onna, Okinawa 904 0495, Japan}
\affiliation[l]{Perimeter Institute for Theoretical Physics,
31 Caroline Street North, Waterloo, ON, N2L 2Y5, Canada}
\affiliation[m]{Department of Mathematics and Statistics, University of New Brunswick, Fredericton, NB, Canada E3B 5A3}
\affiliation[n]{Departamento de F\'{i}sica Te\'{o}rica, Facultad de Ciencias F\'{i}sicas, Universidad Complutense de Madrid, Plaza de las Ciencias 1, 28040 Madrid, Spain, EU}
\affiliation[o]{Department of Physics, Shanghai University, 99 Shangda Rd, 200444, Shanghai, P.R. China}
\affiliation[p]{Scuola Internazionale Superiore di Studi Avanzati (SISSA), Via Bonomea 265,
34134 Trieste, Italy and INFN, Sezione di Trieste, Trieste, Italy}
\affiliation[q]{Instituto de F\'isica, Universidade Federal Fluminense, Campus da Praia Vermelha, Av. Litor\^anea s/n, 24210-346, Niter\'oi, RJ, Brazil}
\affiliation[r]{Theoretisch-Physikalisches Institut, Friedrich-Schiller-Universit\"{a}t Jena, Max-Wien-Platz 1, 07743 Jena, Germany, EU}
\affiliation[s]{Theoretical Particle Physics and Cosmology Group, \, Physics \, Department, King's College London, University of London, Strand, London WC2R2LS, UK}
\affiliation[t]{Mathematical Institute, University of Münster, Einsteinstr. 62, 48149 Münster, Germany}

\date{\today}

\begin{abstract}
{
This collection of perspective pieces captures recent advancements and reflections from a dynamic research community dedicated to bridging quantum gravity, hydrodynamics, and emergent cosmology. It explores four key research areas: (a) the interplay between hydrodynamics and cosmology, including analog gravity systems; (b) phase transitions, continuum limits and emergent geometry in quantum gravity; (c) relational perspectives in gravity and quantum gravity; and (d) the emergence of cosmological models rooted in quantum gravity frameworks. Each contribution presents the distinct perspectives of its respective authors. Additionally, the introduction by the editors proposes an integrative view, suggesting how these thematic units could serve as foundational pillars for a novel theoretical cosmology framework termed \enquote{hydrodynamics on superspace}.
}
\end{abstract}

\maketitle

\newpage

\section{Introduction}\label{sec:introduction}

In recent decades, cosmology has transitioned into an era of high-precision data, enabling extremely accurate measurements of the parameters governing the Lambda-CDM model, as strikingly demonstrated by the PLANCK mission~\cite{Planck:2018vyg}. Its findings provide an exceptionally detailed narrative of the universe’s history, which, however, still reveals significant gaps. In the early universe, General Relativity implies the unresolved puzzle of the Big Bang singularity~\cite{hawking2023large}, alongside questions about the origins of structure formation~\cite{Mukhanov:1990me}. In the late universe, the phenomenon of dark energy—responsible for the observed accelerated expansion—remains only partially explained, with the cosmological constant acting as a phenomenological placeholder without a comprehensive theoretical foundation~\cite{Li:2012dt}. Numerous theoretical models grounded in semi-classical physics exist but they remain fundamentally incomplete. Scenarios for the early universe, such as inflation~\cite{Achucarro:2022qrl}, bouncing cosmologies~\cite{Brandenberger:2016vhg}, and emergent universe models (e.g., string gas cosmology~\cite{Brandenberger:2023ver}), rely on assumptions about the initial state of the universe or invoke new physics to address the singularity, yet lack rigorous control over them. Clearly, a quantum gravity (QG) theory is essential to complete, modify, or potentially replace these cosmological models, providing a coherent framework for both the early and current accelerated expansion of the universe~\cite{Brax:2017idh}.

For QG effects to substantively inform cosmology, they must be embedded in rigorous QG formalisms. The field of quantum gravity is diverse, with each approach offering unique insights into fundamental aspects of the QG problem~\cite{deBoer:2022zka}. Although no approach is yet complete~\cite{Oriti:2009zz,Ashtekar:2014ife,Bambi:2023jiz}, they collectively advance our understanding of how traditional notions of space and time might transform at fundamental scales, potentially replacing the smooth fields of semi-classical physics. Nevertheless, achieving a robust, predictive link between microscopic QG structures and macroscopic, relativistic dynamics remains a challenge. Different QG frameworks exhibit varying degrees of predictive limitations, though this does not negate their potential observational implications~\cite{Barrau:2017tcd}. This has led to the use of phenomenological models, symmetry-reduced toy systems, and QG-inspired modifications at semi-classical levels. While these have yielded interesting results, advancing further certainly requires breaking new ground that can influence both QG and cosmology through interdisciplinary efforts.

The editors of this collection of perspective pieces \textit{Quantum Gravity, Hydrodynamics and Emergent Cosmology}\footnote{This collection is based on the workshop with the same title held on December 8–9, 2022, at the University of Munich (LMU). It was organized by Jibril Ben Achour, Christophe Goeller, Luca Marchetti, Daniele Oriti and Andreas Pithis who are also the editors of the present work.} propose that a novel theoretical cosmology framework, representing an approximate regime of QG, may be grounded in the framework \textit{hydrodynamics on superspace}~\cite{Oriti:2024elx}. This framework posits non-linear dynamics on the configuration space of spacetime fields \enquote{at a spatial point}, with cosmological observables represented as hydrodynamic averages satisfying corresponding equations. Importantly, this formalism should be regarded as a coarse-grained representation of QG, not confined to any specific QG approach but rather adapted to cosmological contexts. Conceptually, this perspective draws on the idea of the emergent nature of spacetime, conceiving the universe as a quantum fluid with cosmological dynamics as a hydrodynamic approximation of the underlying dynamics of \enquote{spacetime constituents}. Thus, this framework serves as a realization of the emergent gravity paradigm~\cite{Linnemann:2017hdo}. Note that the dynamics are not defined on a conventional spacetime manifold but on a field configuration space therein. Spacetime, then, is only recovered as an approximate structure when certain fields serving as rods and clock are treated as physical reference frames, following the relationalist approach widely employed in gravity and QG~\cite{Hoehn:2019fsy}.

Support for this view is twofold. First, recent research has revealed hidden dynamical symmetries within certain relativistic cosmological models that operate in field space~\cite{BenAchour:2022fif,Geiller:2022baq}. This discovery suggests a potential correspondence between cosmology and quantum fluids based on shared symmetries, extending the classical (mean-field) level to a quantum domain, see for instance Refs.~\cite{Lidsey:2013osa,10.1063/1.3429611}.

Second, progress in Tensorial Group Field Theory (TGFT)~\cite{Freidel:2005qe,Oriti:2011jm,Carrozza:2016vsq} connects the hydrodynamics of TGFT condensates to cosmological dynamics. In TGFTs, which are higher-dimensional generalizations of matrix models one incorporates also additional quantum geometric data which enable the exploration of continuum limits and quantum dynamics via Renormalization Group (RG) and mean-field methods. The hydrodynamic equations of TGFT condensates generalize quantum cosmology’s dynamics~\cite{Gielen:2016dss,Oriti:2016acw,Pithis:2019tvp} and have a classical Friedmann limit, with QG corrections producing a quantum bounce instead of a singularity~\cite{Oriti:2016qtz,Marchetti:2020umh,Jercher:2021bie}. This framework also offers potential mechanisms for dark energy~\cite{Oriti:2021rvm,Jercher:2021bie} and new dynamics for cosmological perturbations~\cite{Marchetti:2021gcv,Jercher:2023nxa,Jercher:2023kfr}. Given that the domain of the TGFT condensate wave function is isomorphic to the superspace of homogeneous continuum geometries (plus matter fields), the vision of hydrodynamics on superspace as a coarse-grained description of QG is explicitly realized in the TGFT formalism.

This vision of hydrodynamics on superspace is likely universal. TGFT already represents a synthesis of multiple QG approaches (lattice quantum gravity, tensor models, loop quantum gravity) but similar frameworks emerge from other independent QG methodologies, indicating a shared foundation. For example, a superspace hydrodynamics model arises in loop quantum cosmology (LQC)~\cite{Banerjee:2011qu} within the separate universe approach~\cite{Bojowald:2012wi} and appears in models using mini-superspace quantization of supergravity’s exceptional symmetries~\cite{Kleinschmidt:2009hv}. A related dynamic is seen in the path integral formalism for covariant quantum gravity~\cite{Giddings:1988wv}, where effective superspace hydrodynamics emerge in the continuum limit~\cite{Ambjorn:2009fm}. While these commonalities are impressive, one should bear in mind that each QG formalism may still lead to unique dynamics, and thus to characteristic observable consequences.

Establishing such an overarching framework should provide robust explanations for long-standing cosmological puzzles, but should certainly also advance the ability to derive predictions for observable QG effects in cosmology. However, it is already well known that directly probing the physical conditions in which significant semi-classical and QG phenomena occur remains notoriously challenging. This difficulty has, in part, driven the pursuit of analog gravity models in condensed matter systems, particularly quantum fluids. In such systems, perturbations of the fluids’ hydrodynamic configurations behave analogously to matter fields on a curved geometry defined by hydrodynamic variables and are thus described by quantum field theory (QFT) in curved spacetime. This concept has fueled a wide range of theoretical studies of analog gravitational phenomena, including exotic effects such as black hole evaporation and cosmological particle production, with some of these phenomena even reproduced experimentally in laboratory settings~\cite{Barcelo:2005fc}. However, this approach faces a fundamental limitation: the analogy fails once the dynamical properties of emergent geometry are considered. Specifically, the hydrodynamic equations governing the emergent metric do not map onto gravitational equations. The analog system is a non-relativistic fluid in flat space with preferred reference frames, in contrast to the background and frame independence required by general relativity (GR). Consequently, traditional analog models are constrained to simulate gravitational systems only at the kinematical level. Nonetheless, the editors of this collection propose that this limitation may be overcome with a paradigm shift in perspective: The hydrodynamic/cosmology mappings described above provide examples that circumvent this issue, demonstrating that gravitational dynamics and the hydrodynamics of quantum fluids can indeed align, with their symmetries matching as well. Therefore, to render the proposed framework testable, one should not recreate the spacetime manifold itself in the lab but superspace—the space of field configurations.

This perspective motivated the collection's structure and the selection of its contributors grouped into four thematic units exploring critical aspects of the superspace hydrodynamics framework: (a) Hydrodynamics/cosmology correspondence and link to analog gravity (contributors: Jibril Ben Achour, Salvatore Butera, and Tobias Haas), (2) Phase transitions, continuum limits, and emergence in QG (contributors: Renata Ferrero, Roberto Percacci, Antonio D. Pereira, Dario Benedetti, Sebastian Steinhaus, and Johannes Thürigen), (3) Relational physics in gravity and QG (contributors: Flaminia Giacomini, Kristina Giesel, Joshua Kirklin, and Philipp Höhn), and (4) Emergent cosmology from QG (contributors: Mairi Sakellariadou, Martin Bojowald, Luca Marchetti, and Robert Brandenberger). The first unit focuses on the vision of cosmology as hydrodynamics on superspace and recent developments in the field of analog gravity with the goal of discussing the change of perspective just described. The second unit centers on the use of mean-field and RG techniques to study via coarse-graining the continuum limit and the emergence of spacetime in different QG approaches. The third unit explores recent advances on the relational strategy in gravity and quantum gravity to construct background-independent and diffeomorphism-invariant observables. Finally, the fourth unit sheds light on the intimate relation between cosmology and quantum gravity, as well as emergence of cosmology from quantum gravity approaches. \textcolor{black}{In particular, a better understanding of this cosmic emergence process goes naturally hand in hand with progress in condensed matter physics, renormalization, phase transitions and relational physics. This progress must not only be technical, but also conceptual, as QG forces us to revisit these aspects of classical and quantum physics from its own perspective, stripping them of all the redundant, spacetime based information. This offers a unique chance of cross-fertilization and interdisciplinarity between QG and other fields in physics, as shown in this  collection.} 

\textcolor{black}{However, it is important to emphasize that the contributions can be viewed as standalone perspectives on their respective topics. It was up to the individual contributors to explain whether and how they their research direction could more directly contribute to the proposed framework. Nevertheless, as outlined in the introductory remarks, they clearly inform the framework's development through their results, as further elaborated in the sectional introductions. } Beyond this, note that the invited contributions by proponents of different QG approaches such as asymptotic safety in quantum gravity, causal dynamical triangulations (CDT), spin foam gravity, loop quantum gravity, TGFT and string theory allow one to map the plurality of the QG landscape and to discuss if and how \textcolor{black}{they relate to this framework.}

That being said, it important to note that the editors' vision of hydrodynamics on superspace as a framework for QG and cosmology may potentially not be shared by the various contributors to this collection whose perspective can be gathered from their individual contributions. Correspondingly, beyond the view of the organizers, this collection naturally reports on recent progress in the field of quantum gravity and fosters an exchange of ideas therein, in general. Finally, this collection is concluded in the final section.

\section{Map between hydrodynamics and cosmology}\label{sec:hydrodynamics}

In this first part of the collection, we shall discuss several new results which suggest not only that the quantum cosmological dynamics can be reformulated as a hydrodynamical flow on the space of field configurations, also called superspace (or its extended version dubbed the \textit{lift} hereafter), but also that this shall open the door to develop experimentally testable analog models for quantum cosmology. 

This vision builds on i) mathematical results in symmetry-reduced general relativity and ii) on a paradigm shift in analog gravity. For this, consider the field of analog gravity at first. The ability to reproduce semi-classical effects from quantum field theory on curved backgrounds has remained for a long time a source of debate, mostly driven by a misleading name, starting from the field itself. Analog gravity is not about gravity per se but concerns a quantum field theory on a curved background. This background does not have to be a fixed spacetime representing part of a solution of general relativity, or any other theory of gravity. The only crucial point is that it has to be curved. In particular, this structure can be any other relevant space that one uses to encode the dynamics of a given field theory.

Stepping aside from the standard gravity-oriented interpretation of analog gravity, one immediately sees that the possibilities are much wider. Consider a field theory defined on a curved space. Its low energy regime will be described by hydrodynamics which describe how the lower moments of the field evolve on this curved background. Moreover, these truncated dynamics will be constrained by the symmetry of the underlying space. Identifying the whole set of symmetries constraining the lowest moments of this fluid description allows one to capture the universal properties of these truncated low energy dynamics.

The quantum dynamics of the homogeneous sector of cosmological geometries can be precisely described in this way. This is the standard mini-superspace approach. The quantum cosmological dynamics are described by field evolving on a curved background which is not the spacetime but the configuration space. This space describes all the cosmological metrics (mapped to a two-dimensional space labelled by the scale factor and a matter field in general) and admits in general a non-trivial curvature. Quantum cosmology is then described as a free complex scalar field theory on this curved background. Therefore, if one reproduces this curved background in the lab using a suitable analog system, such as a Bose-Einstein condensate, one could in principle contemplate the dynamics of the wave function of the universe as described in standard quantum cosmology. Nevertheless, in order to build such a quantum universe in the lab, several key notions have to be properly understood and controlled, starting from the dictionary between the observables of the fluid system and the quantum cosmological observables which are by construction relational (e.g. expectation values of the volume w.r.t the reference scalar field for example, see also Section~\ref{sec:relationalism} for further information on the relationalist perspective and Section~\ref{sec:marchetti} for an explicit discussion of this matter in the context of the TGFT condensate cosmology approach), the translation of the meaning of the boundary conditions in the experimental in terms of cosmology, a clear understanding of the gauge symmetry in the cosmological set-up (choice of clock), etc.. 

Nevertheless, if this picture is successfully realized and these various obstacles overcome, it will provide a fascinating perspective for both quantum gravity and analog gravity. On the one hand, we shall have access to experimentally controlled systems mimicking the very early universe, which can be tuned and reproduced at will. On the other hand, it will open the door to bypass a key limitation of analog gravity, namely that the curved background reproduced so far in any system does not contain any dynamical information on the system. By giving up the focus on reproducing the spacetime structure but instead focusing on auxiliary spaces such as the space of field configurations, analog gravity will emancipate itself from its current status \textcolor{black}{to be constrained to simulate gravitational systems only at the kinematical level,} with potential new phenomena to be observed in the lab.

In the following, we discuss three interconnected pieces which advocate for this new point of view, carefully presenting the remaining gaps to be filled before achieving a consistent realization of the paradigm shift. \textcolor{black}{The first contribution, Section~\ref{sec:benachour}, summarizes the various recent results regarding the shared symmetries between hydrodynamics and symmetry-reduced gravitational superspaces, relevant both for cosmology but also for black holes mechanics. This shall facilitate the construction of a dictionary between (quantum) fluids and quantum symmetry reduced gravitational systems. This dictionary suggests that the background configuration of the hydrodynamic flow should be understood as representing not a given spacetime geometry, but an auxiliary space encoding the space of field configurations and, in particular, the dynamical symmetries of the gravitational system. This key point opens the door for analog models to emancipate themselves from their gravity-oriented interpretation and to encode dynamical information on the system of interest at the level of the background. This idea naturally leads to interpreting (quantum) cosmology as a hydrodynamical flow on the space of field configurations. The second contribution, Section~\ref{sec:butera}, reports on recent progress in analog gravity to take into account backreaction effects of quantum fluctuations onto the background geometry. In a sense, this effort goes in the same direction as it aims at encoding effects of the dynamics on the background geometry, unfreezing it from its current kinematical status. Finally, the third contribution, Section~\ref{sec:haas}, reviews recent developments in analog gravity to set up suitable analog models tailored to cosmological settings, in particular to simulate a Friedmann-Lema\^{i}tre-Roberston-Walker geometry and the behavior of quantum fields on it in the lab. Preliminary results aiming at providing an analog BEC model for quantum cosmology and reproducing the WDW dynamics have shown that the geometry of the field space takes a form very close to a cosmological spacetime geometry. It could very well be the case that the current efforts devoted to realizing an analog model for expanding spacetime geometries find another use when constructing such analog models for quantum cosmology.\footnote{For work complementing the content of Sections~\ref{sec:butera} and~\ref{sec:haas}, we refer the reader to Refs.~~\cite{Schutzhold:2005ex,Fischer:2005iy,Baak:2022hum,Pal:2024qno} and~\cite{Tian:2017cda,Ribeiro:2022gln,Cha:2016esj}, respectively.}} These recent progresses underline fundamental questions which are relevant to construct and progress towards the paradigm shift advocated in this manuscript. 

\subsection{Quantum cosmology as hydrodynamics in field space\\ \textit{by Jibril Ben Achour}}\label{sec:benachour}

When investigating the quantum description of the gravitational field, one faces two key challenges: i) identifying the fundamental nature of the microscopic degrees of freedom of the geometry, and ii) understanding the emergence of classical geometries from such microscopic description. That classical gravitational systems can be interpreted as many body systems built from this yet unknown microscopic description of the spacetime fabric is rooted in the fact that the gravitational dynamics admits a thermodynamical interpretation, allowing one to equip black hole but also cosmological horizons (and actually any causal diamond) with an entropy and a temperature.  While there might be different ways to encode the microscopic degrees of freedom depending on the chosen model or theory, one could expect that their dynamics, and thus the emergence of the spacetime geometry in the continuum hydrodynamical approximation, to be governed by universal symmetries. In this contribution, it is reviewed how such universal symmetry emerge in cosmology and how they allow one to recast the Wheeler-DeWitt (WDW) dynamics of quantum cosmology as the hydrodynamical regime of a many-body system. This opens the perspective to interprete quantum cosmology as a type of hydrodynamics on the gravitational mini-superspace. 

A first hint of such symmetries relating the cosmological dynamics and the hydrodynamical regime of a many-body system was presented by Lidsey in Ref.~\cite{Lidsey:2013osa}. To start off, consider a spherically symmetric Bose-Einstein condensate in 2+1 dimension described by the appropriate Gross-Pitaevskii equation. One can show that the first moments of the wave function describing respectively its norm, its width, the radial momentum and energy of the condensate,  form a closed system of equations. Then, provided one identifies the width of the wave function with the scale factor, the radial momentum with the Hubble factor, one can show that this system of equations describing the condensate exactly reproduces the Friedmann equations. While this was initially done for the homogeneous and isotropic universe filled with a scalar field, this duality can be generalized to more complex cosmological systems among which one also has inflationary dynamics~\cite{Lidsey:2018byv}. This surprising duality remains a curiosity and no fundamental nor systematic argument was given for its explanation so far. Yet it suggested a surprising but concrete link between classical cosmology and the hydrodynamical regime of a quantum many-body system. Could this mapping, which holds at the level of the equations of motion, be more fundamental? 

The answer is in the affirmative and came from an independent research direction focusing on the quantization of symmetry-reduced gravitational systems~\cite{BenAchour:2019ufa, Achour:2021lqq, Achour:2021dtj, Geiller:2022baq}. 
To fix the idea, consider a region of spacetime filled with matter and let us assume that the gravitational and matter field are homogeneous. Such region can be understood as a cosmological patch whose dynamics reduces to a pure expansion or contraction induced by the matter. After such drastic symmetry reduction, the full diffeomorphism gauge symmetry reduces to a residual gauge invariance under time-reparametrization. Yet, it was shown in~\cite{BenAchour:2019ufa, Achour:2021lqq} that on top of this gauge invariance, the system still enjoys an additional physical SL$(2,\mathbb{R})$ symmetry. At the phase space level, this symmetry can be recognised from the $\mathfrak{sl}(2,\mathbb{R})$ algebra formed by the 3d volume of the region, the Hamiltonian constraint and the trace of the extrinsic curvature, known as the CVH algebra. This structure and its generalizations studied in~\cite{BenAchour:2020njq, BenAchour:2020ewm, BenAchour:2020xif, Achour:2021dtj, Geiller:2022baq} were first used as a symmetry-based argument to perform an ambiguity-free canonical quantization of cosmological (and black hole) mechanics~\cite{BenAchour:2019ywl, Sartini:2021ktb}. Yet, a clear understanding of these symmetries was only achieved recently. Indeed, these symmetries are not spacetime symmetries but act in field space. A powerful tool to identify and analyze these hidden symmetries is the so called Einsenhar-Duval lift~\cite{Cariglia:2016oft}. This systematic approach was applied to both the cosmological and black hole symmetry-reduced models of GR in~\cite{BenAchour:2022fif}. As expected, the dynamics of any homogeneous gravitational system can be described as the one of a massless or massive test particle propagating in an auxiliary geometry known as the lift (also called the extended superspace). Therefore, the conformal isometries of the lift metric naturally become a physical symmetry of the reduced action describing the homogeneous gravitational system. How does this relate to hydrodynamics and to many-body systems?

The interesting outcome is that following this strategy, one can show that both cosmological and black hole mechanics are invariant under the Schr\"{o}dinger group~\cite{BenAchour:2022fif, BenAchour:2023dgj}, a symmetry shared by  the compressible Navier-Stokes equation~\cite{Horvathy:2009kz} but also by a whole family of non-linear Schr\"{o}dinger (NLS) equations describing Bose-Einstein condensates in various dimensions~\cite{Niederer:1972zz, Ghosh:2001an, Kolomeisky:2000zz}. The key point is that the lift associated to these symmetry-reduced gravitational systems admits a Schr\"{o}dinger-like group as its conformal isometries. It might sound surprising at first that this symmetry, which is a property of massive non-relativistic systems, shows up in gravity. However, there is one key difference with the standard Schr\"{o}dinger symmetry: Because the lift has a signature $(+,+,-,-)$ for the simplest homogeneous gravitational systems studied in~\cite{BenAchour:2022fif}, the field space admits two time directions. This changes the compact rotation generator of the  Schr\"{o}dinger algebra to a non-compact boost, hence the Schr\"{o}dinger-like structure. 

Despite this fact, the existence of this shared symmetry between NLS equations and the homogeneous sector of GR allows one to extend the duality proposed by Lidsey in~\cite{Lidsey:2013osa} beyond the level of the equation of motion, up to the Lagrangian symmetries of the systems. It provides a first hint to recast quantum cosmology and quantum black hole mechanics as the hydrodynamical description of a many-body system. \textit{The crucial point is that the associated fluid lives not in spacetime but in field space.} 

This point of view resonates with different approaches of quantum gravity which aim at recovering the cosmological dynamics as an hydrodynamical approximation through coarse-graining of pre-geometric building blocks. Important efforts have been developed in this direction in the group field theory (GFT) approach to quantum gravity where the bouncing cosmological dynamics with the right semi-classical limit was recovered following this scheme~\cite{Gielen:2013naa, Oriti:2016qtz,Pithis:2019tvp,Jercher:2021bie}\textcolor{black}{, see in particular Section~\ref{sec:marchetti} for a lightning review of the TGFT condensate cosmology approach. Now, such derivation requires several layers of approximations which need to be controlled in order to fully develop a consistent emergent cosmology from these GFT models. While efforts are devoted to further refine and characterize this coarse-grained perspective corresponding to a mean-field approximation~\cite{Marchetti:2020xvf,Marchetti:2022igl,Marchetti:2022nrf,Dekhil:2024ssa,Dekhil:2024djp}, it is interesting to see that one can extract directly from the IR regime symmetry-based-arguments which reinforce the picture of quantum cosmology as emerging from an hydrodynamical description in field space of a many-body system. These developments demonstrate that TGFT condensate cosmology is a specific instantiation of the \textit{hydrodynamics on superspace
framework}, featured in the general introduction to this collection. We refer the reader to Ref.~\cite{Oriti:2024qav} for a detailed presentation of this new framework.}

To conclude, it should be stressed that this result also suggests that the standard WDW dynamics used in quantum cosmology (but also in quantum black hole mechanics) could be generalized by non-linear interaction of the wave function of the geometry just as the free Schr\"{o}dinger equation can be extended to Gross-Pitaevskii equation when describing condensates. Indeed, the Schr\"{o}dinger invariance allows one to uniquely select non-linear corrections which preserves this symmetry, giving rise to a one parameter family of non-linear WDW equations~\cite{BenAchour:2023dgj}. Therefore, it would be useful to explore what could be the phenomenological consequences of this new non-linear extension of the WDW dynamics in quantum cosmology.  Nevertheless, more work is needed to understand how this physical symmetry can be contemplated and exploited without fixing the gauge. To that end, it would be useful to extend the Einseinhart-Duval lift to fully time-reparametrization mechanical systems, by understanding how the lapse can be included in this geometric framework.  

Yet, already at this level, the present finding opens the fascinating possibility to identify suitable analog models of quantum cosmology, where the background would play the role of the lift and the perturbation would represent the wave-function of the universe. If such analog model can be realized, this would provide a first approach to go beyond the current state-of-the-art in analog gravity and reproduce the full dynamics of a given (quantum) gravitational system in the lab. Efforts in this direction are currently in progress. 

Finally, an interesting perspective would be to connect the above findings to the non-relativistic holography and its gravity/cold atoms duality developed in~\cite{Son:2008ye, Taylor:2008tg} where the Schr\"{o}dinger symmetry plays a key role. Another interesting question is whether the two time directions present in the gravitational lift could allow non-relativistic (invariant under the Schr\"{o}dinger group) and ultra-relativistic (invariant under the Carrolian group) regimes to coexist.

\subsection{Exploring back-reaction with analog gravity models\\ \textit{by Salvatore Butera}}\label{sec:butera}

It is widely believed that general relativity, and the description of matter in terms of quantum fields living on a smooth spacetime manifold, emerge as the low-energy effective description of a more fundamental microscopic theory~\cite{hu2005can}. Such a hydrodynamic description is expected to breakdown at the higher energies probing the Planck scale physics, at which point the concept of spacetime itself looses meaning. If we accept this picture, it is not a surprise that fields experiencing a non-trivial effective spacetime appear as the hydrodynamic (long-wavelength) description of condensed-matter systems, whose microscopic dynamics has nothing to do with gravity and its hypothetical Planck scale physics. Since Unruh's seminal paper~\cite{Unruh-Analog-1981}, these \emph{analog models} of gravity~\cite{Barcelo-2011} have become a powerful platform for theoretically investigating effects of quantum fields in curved spacetime, and provided experimentalists with physical systems where these effects can be observed in table-top experiments. A most celebrated achievement in the field is the pioneering observation of the Hawking radiation, emanating from a sonic black hole~\cite{Steinhauer2016,Steinhauer2019,Steinhauer2021} implemented in trans-sonically flowing Bose-Einstein condensates (BECs) of ultra-cold atoms~\cite{Carusotto2008,Lahav2010}. The analog of cosmological particle creation has also been investigated theoretically~\cite{Fedichev2004,Uhlmann2005,Jain2007,Butera2021} and experimentally. Specifically, in the case of an expanding system, both with BECs ~\cite{Eckel2018,viermann2022} and quantum fluids of light~\cite{steinhauer2021-Light}, and in the case of an analog configuration of cosmological preheating, both with BECs~\cite{Gabri-PreHeat-2022} and classical fluids~\cite{Silke-PreHeat-2022}. Furthermore, the superradiant scattering from rotating sonic black holes has been observed in a water tank exhibiting a draining vortex flow configuration~\cite{Silke-SuperRad-2017}.

These achievements demonstrate the capability of analog systems to describe test-field effects of (non-interacting) quantum fields in curved spacetime~\cite{birrell1984quantum}. The question naturally arises whether the field of analog gravity can be extended to simulate back-reaction effects~\cite{Balbinot2005}, potentially shedding light on the intricate physics that arise from the interplay between quantum fields and spacetime. Clarifying this interplay is of fundamental importance in gravity, to gain a deeper understanding of those physical configurations in which gravitational interactions are strong and quantum effects are important. Pioneering studies in the field anticipated that quantum effects may have had a profound influence in driving the evolution of the early universe towards the present stage~\cite{hu2020semiclassical}. Related to this, black holes are expected to evaporate and eventually disappear due to the emitted Hawking radiation~\cite{Hawking1975}, in which case we believe that quantum back-reaction effects may be involved in the resolution of the so-called information paradox~\cite{Maldacena-InfoParadox-2020}.

The difficulty we face in attempting to investigate back-reaction with analog models lies in the different non-linear dynamics of these systems compared to what prescribed by general relativity. Moreover, there is no reason to expect that condensed-matter microscopic physics can reproduce the physics of spacetime at the Planck scale. In other words, the following question arises: What can we learn about back-reaction if the fundamental equations governing gravity and analog systems are different? One possible answer draws inspiration from the effective field theory approach~\cite{EFT}, and relies on the fact that mesoscopic back-reaction effects such as fluctuation, dissipation, and decoherence emerge from coarse-graining the microscopic physics and are thus expected to be universal. Following this phenomenological approach, analog systems can be envisioned as platforms to investigate qualitative features of the back-reaction and clarify related fundamental processes.

Remarkably, first experiments investigating classical back-reaction effects have been recently reported~\cite{Patrick:PRL21}. Additionally, pioneering theoretical works appeared, investigating back-reaction physics of quantum fluctuations in a cold atoms analog of early universe pre-heating~\cite{Robertson-PRD-PreHeatAn-2019,butera-preheating-2022}. These works focused on non-perturbative effects of particle production onto the inflaton dynamics, demonstrating the occurrence of quantum friction, spatial dephasing of the inflaton oscillations and the consequent fragmentation~\cite{butera-preheating-2022}. Remarkably, these works gave evidence of the crucial role of quantum fluctuations in the dynamics of the analog inflaton field beyond the semiclassical level~\cite{Pla:PRD2021}.
These early results point in the direction of extending the analog gravity program beyond the standard test field level of quantum field theory in curved spacetime, opening the way to using analog models as a quantum simulation platform for a novel set of problems of current interest in cosmology. The next challenge facing the analog gravity community is to extend these studies to the more complex configuration of an analog black hole, with the aim of clarifying quantum back-reaction effects induced by Hawking emission on the horizon~\cite{Hu_BH_BR}. Building upon recent results obtained with simpler models~\cite{Trilinear-theory}, it is expected that the tripartite entanglement between the black hole's vibrational modes and the Hawking and partner modes on either side of the horizon, may provide a mechanism for communication between the black hole's interior and exterior. This potential information leakage could be revealed by observing deviations from the thermal spectrum of analog Hawking emissions in future experiments, potentially offering a resolution the long-standing information paradox.

\subsection{A Universe in Heidelberg\\ \textit{by Tobias Haas}}\label{sec:haas}

When considering quantum fields in \textit{curved} spacetimes, seemingly simple concepts such as the vacuum or the notion of particles become observer-dependent~\cite{Birrell1982,Mukhanov2007,Weinberg2008}. For instance, an observer far away from the boundary of a black hole perceives a thermal spectrum~\cite{Hawking1975}, while an observer in an expanding universe experiences the production of particles~\cite{Parker1969}. 

Despite the wide acceptance of these theoretical predictions, detecting the corresponding phenomena in the night sky remains an open challenge. Over the past two decades, substantial progress has been made by focusing on analogies of such effects instead. These so-called analog gravity models emerge generically for the perturbations of barotropic, inviscid, and irrotational fluids in form of an acoustic metric~\cite{Unruh1981,Visser1998,Visser2002,Volovik2009,Barcelo2011}
\begin{equation}
    \mathrm{d}s^2 \propto -(c^2-v^2) \mathrm{d}t^2 - 2 \boldsymbol{v} \mathrm{d}\boldsymbol{x} \mathrm{d}t + \mathrm{d}\boldsymbol{x}^2,
\end{equation}
where $\boldsymbol{v}$ denotes the background fluid velocity and $c$ the speed of sound, which corresponds to the causal speed in the emergent spacetime.


This approach did not only add another connection between hydrodynamics and gravity but also paved the ground for accessing the effects of quantum fields in curved spacetime in the lab. In this way, theoretical predictions on analogs of black holes~\cite{Unruh1981,Unruh1995,Visser1998,Garay2000,Garay2001,Novello2002,Barcelo2003,Balbinot2005,Schuetzhold2010,Fabbri2021}, Unruh radiation~\cite{Leonhardt2018}, and cosmological particle production~\cite{Barcelo2003b,Barcelo2003c,Fedichev2003,Fedichev2004,Fischer2004,Fischer2004b,Uhlmann2005,Calzetta2005,Jain2007,Weinfurtner2009,Prain2010,Bilic2013} have been realized in classical~\cite{Philbin2008,Weinfurtner2011,Patrick2021} as well as quantum~\cite{Carusotto2008,Lahav2010,Horstmann2010,Steinhauer2014,Steinhauer2016,Eckel2018,Hu2019,MunozDeNova2019,Wittemer2019,Jacquet2020,Gooding2020,Banik2021,Kolobov2021,Steinhauer2022} fluids. 

In 2022, a Bose-Einstein condensate, which exhibits the properties of a superfluid, was set up in Heidelberg to serve as a quantum field simulator for the dynamics of a relativistic scalar field in an expanding Friedmann-Lema\^{i}tre-Robertson-Walker universe of arbitrary spatial curvature~\cite{Vierman2022}. While building upon the mentioned advances in the field and the theoretical works~\cite{Tolosa-Simeon2022,Sanchez-Kuntz2022}, this study comprises not only the first \textit{strict} one-to-one correspondence between acoustic excitations in a Bose-Einstein condensate and a scalar field in an expanding universe but also the first observation of analog cosmological particle production in agreement with cosmological predictions.

Based on these recent advances, we discuss a few interesting research directions to further sharpen our understanding of quantum field theoretic effects in cosmological settings. When sticking to the standard analog gravity paradigm, that is, a classical spacetime geometry fully determined by experimental parameters, quantum effects solely arise for the scalar field experiencing the emergent spacetime. In this context, straightforward upgrades of the scenarios described in Ref.~\cite{Vierman2022} would be to study not only expanding but also contracting or even cyclic universes. Similarly, more rapid expansions, such as for example of de Sitter type, in which case the space expands on the order of 10 or more e-folds, should be investigated. 

A more challenging question regards the quantum nature of the particle production process. Theoretical studies indicate that particles are produced in pairs with entanglement across the two constituents~\cite{Bruschi2013,Robertson2017a,Robertson2017b,Chen2021}. Although spatial correlations have been observed in Ref.~\cite{Vierman2022}, finding efficient methods for certifying and measuring \textit{quantum} correlations in cosmological settings remains an important open problem.

It is an even more ambitious goal to formulate a hydrodynamical analogy for models of \textit{quantum gravity}. First steps in this direction have been made by including backreaction effects of the particle production process, which sources the energy-momentum tensor (or at least expectation values thereof), onto the dynamics of spacetime~\cite{Hu2020,Patrick2021,Butera2022}, see Section~\ref{sec:butera}.

\textcolor{black}{Following this perspective’s spirit of "hydrodynamics on superspace" and given the deep connection between hydrodynamic and cosmological models in their symmetries~\cite{BenAchour2022}, see also Section~\ref{sec:benachour}, one might consider the fascinating possibility of hydrodynamic analogies for \textit{quantum cosmology} models, see the first contribution for details. The equation of motion for the wave function of the universe, the Wheeler-DeWitt (WDW) equation~\cite{Hartle1983,Kiefer2022}, which might be quantum simulated by the dynamics of acoustic excitations in ultracold atom experiments similar to~\cite{Vierman2022} - when viewed through the lens of the hitherto discussed superspace theory. In this way, predictions of quantum cosmology models could ultimately be rendered accessible to experimental tests~\cite{Carney2019} and analog gravity experiments would become the window into one of the most concealed pieces of our world: the quantum nature of spacetime.}

\section{Phase transitions, continuum limits and emergence in quantum gravity}\label{sec:phasetransitions}

The quantization of gravity has posed a fundamental challenge in physics for many decades. While the path integral method applied to the Einstein-Hilbert action provides a viable framework within effective field theory in the weak-field limit~\cite{Donoghue:2012zc}, one encounters perturbative non-renormalizability, leading to a loss of predictability~\cite{tHooft:1974toh,Goroff:1985th}.

An alternative approach to quantizing gravity within the continuum path integral framework is explored through the asymptotic-safety program~\cite{Bonanno:2020bil}, as detailed in Section~\ref{sec:ferrero} of this collection. This approach is based on the hypothesis of an interacting fixed point in the Renormalization Group (RG) flow for gravity in the ultraviolet (UV) regime~\cite{Weinberg:1976xy,Weinberg:1980gg,Reuter:1996cp}, which bypasses the aforementioned difficulties associated with perturbative quantization. Should such a fixed point indeed exist, it would provide a well-defined continuum limit for the path integral. The functional Renormalization Group (FRG)~\cite{Berges:2000ew,Delamotte:2007pf,Dupuis:2020fhh} is a commonly employed method for investigating this hypothesis, effectively implementing the Wilsonian idea of coarse-graining. Many explicit computations within truncated RG flows support the existence of such a fixed point~\cite{Percacci:2017fkn,Reuter:2019byg,Bonanno:2020bil,Reichert:2020mja}.

Another perspective on evaluating the path integral over geometries, potentially including a sum over topologies, involves reformulating the continuum action in a discretized form (potentially using more suitable variables) and then summing over discrete triangulations. In this framework, continuous spacetime geometry is expected to emerge through a phase transition. However, controlling the continuum limit and extracting effective gravitational dynamics remains a central challenge for approaches based on discrete structures. This class of methods includes matrix~\cite{DiFrancesco:1993cyw} and tensor models~\cite{Gurau:2016cjo,GurauBook,Eichhorn:2018phj,Gurau:2024nzv}, group field theory (GFT)~\cite{Freidel:2005qe,Oriti:2011jm,Carrozza:2013oiy}, spin foam gravity~\cite{Perez:2003vx,Perez:2012wv}, and both Euclidean and Causal Dynamical Triangulation approaches (EDT and CDT)~\cite{Ambjorn:2012jv,Ambjorn:2013tki,jordan2013globally,Loll:2019rdj}, with the latter three discussed further in this section. Notably, in GFT models exhibiting asymptotic freedom~\cite{BenGeloun:2012pu,BenGeloun:2012yk} and asymptotic safety~\cite{Carrozza:2014rya,Carrozza:2016tih} have been obtained. The FRG, adapted recently to discrete quantum gravity frameworks~\cite{Eichhorn:2013isa,Eichhorn:2014xaa,Benedetti:2015et,Benedetti:2016db,BenGeloun:2016kw,Eichhorn:2017xhy,Eichhorn:2018phj,BenGeloun:2018ekd,Eichhorn:2019hsa,Castro:2020dzt,Eichhorn:2020sla,Pithis:2020kio,Geloun:2023ray}, has also proven here to be a powerful tool for mapping the phase structure of various models, as demonstrated in~\cite{Benedetti:2015et,Benedetti:2016db,BenGeloun:2016kw,BenGeloun:2018ekd,Pithis:2020kio,Geloun:2023ray} and reviewed in~\cite{Carrozza:2016vsq,Carrozza:2024gnh}. Additionally, mean-field theory techniques have recently been applied in this context~\cite{Pithis:2018eaq,Pithis:2019mlv,Marchetti:2020xvf,Oriti:2021oux,Marchetti:2022igl,Marchetti:2022nrf,Dekhil:2024ssa,Dekhil:2024djp}, see Section~\ref{sec:thuerigen} of this article, to demonstrate evidence of a viable continuum gravitational regime for increasingly realistic GFT models of $4$-dimensional Lorentzian quantum gravity, with~\cite{Marchetti:2022nrf} providing notable results.

Similarly, RG techniques are being developed to investigate the critical properties and the phase diagram of spin foam models~\cite{Dittrich:2014ala,Delcamp:2016dqo,Bahr:2016hwc,Steinhaus:2018aav,Bahr:2018gwf,Steinhaus:2020lgb,Asante:2022dnj}, discussed specifically in Section~\ref{sec:steinhaus} of this collection. These efforts can be compared to developments in the Euclidean and Causal Dynamical Triangulation approaches (EDT and CDT)~\cite{Ambjorn:2022naa}, where significant progress has been made in understanding the phase diagram. In CDT, for instance, a phase with physically relevant, extended macroscopic geometries has been identified~\cite{Ambjorn:2004qm,Ambjorn:2005db,Ambjorn:2007jv,Ambjorn:2012jv}, bounded by a second-order phase transition that permits a well-defined continuum limit~\cite{Ambjorn:2011cg,Ambjorn:2012ij,Ambjorn:2016mnn}, as further explored in Section~\ref{sec:benedetti} of this collection.

\textcolor{black}{The goal of this section is to present recent advances and to highlight outstanding challenges within these approaches. In particular, each contribution can be seen as a compact standalone introduction to its respective topic. Next to this, it also establishes connections to Section~\ref{sec:relationalism} on relationalism and Section~\ref{sec:cosmology} on deriving effective cosmology from quantum gravity. In view of the \textit{hydrodynamics on superspace} framework advertised by the editors, this section discusses in part the topic of emergence in quantum gravity from discrete structures.  In particular, Sections~\ref{sec:benedetti} and~\ref{sec:thuerigen} advertise the use of mean-field techniques in the CDT and TGFT approaches to this end. While it is left to future investigations, how the results presented in Section~\ref{sec:benedetti} could fit in detail into this novel framework, the results on phase transitions and condensate states in TGFT, displayed by Section~\ref{sec:thuerigen}, directly support this approach as an instantiation of it. This is because the mean-field hydrodynamics of such states can be mapped to an effective continuum cosmological dynamics~\cite{Gielen:2013kla,Oriti:2016qtz,Pithis:2019tvp,Jercher:2021bie}, see especially Section~\ref{sec:marchetti} for an elaboration of this matter. Also, the cosmological dynamics of effective spin foam models resulting from coarse-graining, see Section~\ref{sec:steinhaus}, may potentially fit into this framework. Clearly, the perspective discussed in Section~\ref{sec:ferrero} on asymptotic safety approach is different in this regard since it is based on fields living on continuum spacetime. However, the investigation of the phase structure therein and in particular the exploration of the continuum limit is a shared challenge among all of these approaches and shared tools, like the renormalization group, are successfully employed to this end. Beyond this, in anticipation of Section~\ref{sec:relationalism} on the relationalist perspective, Section~\ref{sec:ferrero} reports on recent efforts to employ relational frames in the asymptotic safety approach which is a common tool in quantum gravity approaches to construct gauge-invariant observables and to deparametrize the full dynamics. This method plays a highly important role in the \textit{hydrodynamics on superspace} framework where spacetime is expected to be recovered as an emergent and approximate structure using it.}

\subsection{Effective dynamics for the spatial volume of 3-dimensional CDT\\ \textit{by Dario Benedetti}}\label{sec:benedetti}

Causal dynamical triangulations (CDT) is an approach to quantum gravity based on the idea of regularizing the path integral over geometries by turning it into a sum over piecewise flat manifolds with a preferred foliation (see~\cite{Ambjorn:2012jv,Loll:2019rdj} for reviews). One of the open questions in CDT is whether such a preferred foliation plays a crucial role in the continuum limit, 
for example leading to a Lorentz-violating theory such as Ho{\v{r}}ava-Lifshitz gravity (see~\cite{Wang:2017brl,Steinwachs:2020jkj} for reviews), or whether it is only part of a convenient regularization structure that disappears in the continuum.

One feature of CDT is the existence of a phase characterized by the emergence of a macroscopic universe, suggestive of a semiclassical behavior at large scales. A spacetime condensation phenomenon underlies the emergence of such a semiclassical universe, 
where the time extension of the condensate, or droplet, is strictly smaller than the total time of the quantum universe. In~\cite{Benedetti:2014dra,Benedetti:2016rwo,Benedetti:2022ots} we introduced an effective model for the dynamics of the spatial volume in $2+1$ dimensions, which can be obtained as a mini-superspace reduction of Ho{\v{r}}ava-Lifshitz gravity, and we have shown that a study of the minima of the action (supported also by Monte-Carlo simulations) leads to a phase diagram with a droplet phase similar to that of CDT, which instead cannot be reproduced with a model based on a mini-superspace reduction of Einstein's gravity theory. Therefore, in view also of previous results ~\cite{Horava:2009if,Benedetti:2009ge,Ambjorn:2010hu,Budd:2011zm,Ambjorn:2013joa}, we conclude that there is strong evidence that the effective field theory associated to the continuum limit of CDT in $1+1$ and $2+1$ dimensions is a foliation-preserving gravity theory,
that is, a theory in the same general class to which Ho{\v{r}}ava-Lifshitz gravity belongs (see~\cite{Benedetti:2022ots} for a detailed argument\textcolor{black}{, where the effective mini-superspace models are interpreted as a mean-field theory of CDT}). 

Notice that such conclusion is not in contradiction with the existence of a generalized version of CDT with a local causality constraint that relaxes the preferred foliation at the level of the lattice~\cite{Jordan:2013iaa,Jordan:2013awa,Loll:2015yaa}, because what matters for the continuum description is the infrared behavior: as a manifestation of universality, any modification of the lattice model, that leads to the same infrared physics, would fit in the same effective field theory description.

The question of identifying the correct space of effective field theories is instead still open in $3+1$ dimensions, where it is  harder to address, but there is currently no clear-cut argument why the status of the foliation should be different in that case.

\subsection{Spin foams, refinement limit and semi-classical physics\\ \textit{by Sebastian Steinhaus}}\label{sec:steinhaus}

Spin foam models~\cite{Perez:2012wv, Engle:2023qsu} are a path integral approach of quantum gravity derived from a gauge theoretic formulation of gravity. The continuous manifold is replaced by a discretization and labelled with group theoretic data to encode the (quantum) geometry~\cite{Baez:1999tk}, e.g. irreducible representations of the symmetry group as areas of surfaces, as in spin network states of loop quantum gravity~\cite{Rovelli:2004tv,Thiemann:2023zjd}. A spin foam model locally assigns amplitudes to these (quantum) geometric building blocks, which are derived from general relativity written as a constrained topological theory~\cite{Plebanski:1977zz}. 

The role of the discretization requires special attention in spin foams~\cite{Dittrich:2008pw}. On the one hand, we can understand it as a regulator, a necessity to define the path integral. As a fiducial object, it has no inherent physical meaning and we must ensure that results are independent of this choice. The goal thus is to remove this regulator in a so-called refinement limit~\cite{Dittrich:2013xwa,Steinhaus:2020lgb,Asante:2022dnj}. On the other hand, spin foam amplitudes can be interpreted as Feynman diagrams of the perturbative expansion of group field theories (GFT)~\cite{Oriti:2006se}, also discussed in Sections~\ref{sec:thuerigen},~\ref{sec:marchetti} and~\ref{sec:sakellariadou} of this collection. GFTs are quantum field theories of spacetime, defined on (several copies of) group manifolds. In this framework, one sums over all possible discretizations and geometries, rendering the theory discretization independent. Still, the renormalizability of different models as quantum field theories has to be checked~\cite{Carrozza:2016vsq}. While both approaches differ in tools used, they provide crucial complementary perspectives. In the following, we focus on the former approach, namely on ``refining spin foams''.

\paragraph{Refinement limit of spin foams.}
A refinement limit in spin foams can be formulated with respect to the boundary of a spin foam~\cite{Dittrich:2013xwa}: given a boundary state, a spin foam model assigns an amplitude to each state in its boundary Hilbert space. If the boundary is bipartite, this is frequently called a transition from an initial to a final boundary state. A discretization independent theory would then imply that results/predictions agree an any discretization of the same physical process. The implied challenge is two-fold as we must find states in different boundary Hilbert spaces describing the same boundary (quantum) geometry and then to assign the dynamics, i.e. amplitudes, for the results to agree for any (bulk) discretization. Realizing this ambitious endeavor essentially implies perfectly representing the dynamics on any discretization, i.e. solving the theory.

Examples of such realizations exist, e.g. topological theories~\cite{ponzano1968semiclassical} or classical Regge gravity~\cite{Regge:1961px} for flat solutions~\cite{Rocek:1982tj}, yet it is difficult to realize it without approximations for interacting theories such as 4d gravity. Instead, the strategy would be to improve the theory~\cite{Bahr:2009qc}, e.g. via coarse graining. To improve spin foams, we must define a concept called embedding maps: these maps relate coarse to fine boundary Hilbert spaces, adding degrees of freedom. This concept exists also for spin network states in form of the Ashtekar-Lewandowski vacuum~\cite{Ashtekar:1993wf} and the Geiller-Dittrich vacuum~\cite{Dittrich:2014wpa}. Thus, these maps help to translate coarse to fine transitions. Conversely, we can use these maps to define a coarse graining flow of amplitudes: Interpreted in the other direction, embedding maps project fine degrees of freedom to effective coarse ones. Then, we define effective coarse amplitudes by acting with embedding maps on fine boundaries and integrating out the fine degrees of freedom. This defines a renormalization group flow (dependent on the choice of embedding maps), from fine to coarse discretizations. Essentially we interpret the discretization as a relative, albeit unphysical, scale and associate different amplitudes to discretizations. To define the refinement limit, we search for fixed points of this flow on a second-order phase transition to define a non-trivial continuum theory. For more elaborate arguments, we refer the reader to recent reviews on this subject~\cite{Steinhaus:2020lgb,Asante:2022dnj}, including discussions of concrete numerical realizations e.g. utilizing tensor network methods~\cite{Dittrich:2014mxa,Delcamp:2016dqo,Cunningham:2020uco} or restricted spin foam models~\cite{Bahr:2015gxa,Bahr:2016hwc}.

Due to the numerical complexity of computing the non-perturbative dynamics, coarse graining 4d spin foam models is challenging. In the following, we will elaborate on how new more efficient, asymptotic methods may be developed to accelerate calculations.

\paragraph{Hybrid representation and semi-classical physics.}
Independent of the concrete renormalization efforts, we must study the non-perturbative dynamics of spin foams on large discretizations; efficient numerical methods are indispensable. In recent years there has been significant progress~\cite{Dona:2019dkf,Gozzini:2021kbt,Dona:2023myv,Steinhaus:2024qov,Asante:2024eft} that opens the road to tackling such calculations for larger discretizations, yet cannot be continued indefinitely due to rapidly growing numerical costs. Instead, the expectation is that beyond a certain scale, in terms of the size of representations, asymptotic methods can complement numerical calculations by providing viable approximations as e.g. convincingly shown for a single vertex amplitude~\cite{Conrady:2008mk,Barrett:2009gg,Barrett:2009mw}.

Hence the task is to find the dominant contributions to the spin foam path integral. Two approaches tackle this question from different directions. Complex critical points~\cite{Han:2021kll} generalize the asymptotic analysis from real to complex critical points, thus e.g. including curved configurations. Effective spin foams~\cite{Asante:2020qpa,Asante:2021zzh} on the other hand postulate a modified area Regge calculus~\cite{Barrett:1997tx,Asante:2018wqy} path integral, in which the geometry is built from non-matching geometric 4-simplices. Non-matching, which allows for non-metricity, is exponentially suppressed by so-called gluing constraints .

The hybrid representation~\cite{Asante:2022lnp} combines these two pictures in full spin foam models. Using several resolutions of identity for coherent states, we can mimic the effective spin foam paradigm. Each 4-simplex is equipped with its independent set of coherent boundary data between which we interpolate. Both numerically and asymptotically, one can show that these interpolations, also called gluing constraints, are peaked on matching geometries between the two 4-simplices. These results strongly resonate with the underlying assumptions of effective spin foams and may help in searching for complex critical points~\cite{Han:2024lti}. Moreover, these methods will be vital to compare to classical results and determine whether we can recover semi-classical physics from spin foams.

\paragraph{Outlook.}
Effective, asymptotic methods may pave new ways towards investigating renormalization and discretization (in)dependence in spin foams. A natural scenario to tackle this question would be cosmological spin foam path integrals: the high degree of symmetry drastically reduces the number of degrees of freedom, the classical dynamics are available both in the discrete~\cite{Hartle:1985wr,CorreiadaSilva:1999cg,Dittrich:2021gww,Jercher:2023csk} and the continuum and time reparametrization invariance exists as a remnant of diffeomorphism symmetry. Interesting results have started to emerge recently~\cite{Dittrich:2023rcr,Han:2024ydv}.

\textcolor{black}{On a more speculative note, the question arises how cosmology may emerge from the full theory before imposing symmetry restrictions. Cosmological dynamics may emerge (on average) by considering coarse-grained observables, like the 3-volume of a spatial slice, and matter fields as relational clocks and rods, revealing similarities and complementary perspectives to the \textit{hydrodynamics on superspace} program.}

\subsection{Three Ways to Mean-Field Group Field Theory\\ \textit{by Johannes Thürigen}}\label{sec:thuerigen}

Tensorial group field theory (TGFT)~\cite{Freidel:2005qe,Oriti:2011jm,Carrozza:2013oiy} is a framework providing a path integral for the geometric degrees of freedom of general relativity on discrete spacetime as well as for an ensemble of such spacetimes themselves, built from discrete building blocks. In this way it allows to connect to the observable smooth spacetime geometry within the realm of continuum random geometries. Importantly, TGFT is closely related to other approaches to quantum gravity, e.g. it offers a reformulation of specific simplicial lattice gravity path integrals~\cite{Bonzom:2009hw,Baratin:2010wi,Baratin:2011tx,Baratin:2011hp,Finocchiaro:2018hks} and provides a completion of the quantum dynamics described in spin foam models~\cite{Rovelli:2011eq,Perez:2012wv,Livine:2024hhc}, which is a covariant counterpart of canonical loop quantum gravity~\cite{Ashtekar:2004eh,Thiemann2007a}. With TGFT being a field theory, a broad range of well established QFT methods are at hand, though their application to TGFT is particularly challenging due to its specific properties of tensor fields interacting in a combinatorially non-local way, the Lie group domain and its geometry, as well as so-called geometricity constraints. 
It is a central question in TGFT, as much as in quantum gravity in general, whether and under what conditions there is an appropriate phase of continuum geometry.

\newcommand{\efd}{{d_{\textrm{eff}}}}
\newcommand{\crd}{{d_{\textrm{crit}}}}
\newcommand{\gd}{{d_{\textsc{g}}}}

The result presented in this contribution is that we can identify such a phase with a non-trivial vacuum that can be described by mean-field theory.
Three properties of TGFT are crucial to this end~\cite{Marchetti:2022nrf}:
(1)~Despite of the combinatorial non-locality, TGFT scales like a local field theory but with some modified, ``effective'' dimension~$\efd$ depending on the combinatorics of the dominant interactions~\cite{Pithis:2018eaq,Pithis:2020sxm,Pithis:2020kio}.
(2)~Adding $d_{\phi}$ scalar matter degrees of freedom to the gravitational ones increases this dimension $\efd \to \efd + d_{\phi}$~\cite{Marchetti:2020xvf, Geloun:2023ray}. 
(3)~The geometry of the Lie group $G$ can lead to a flowing effective dimension $\efd(k)$ depending on a scale~$k$, similar to local QFT~\cite{Benedetti:2014gja}. In particular, on a compact domain it vanishes in the IR, $\efd(k)\to 0$ for $k\to 0$~\cite{Pithis:2020kio,Pithis:2020sxm,Geloun:2023ray}, while on group with hyperbolic geometry $\efd(k)\to\infty$~\cite{Marchetti:2022igl, Marchetti:2022nrf}.

Mean-field theory is applicable if $\efd > \crd$. 
This works even for compact group if there are sufficiently many matter degrees of freedom adding to $\efd$. 
On groups with hyperbolic geometry, as natural when there is some Lorentzian structure in a TGFT model as in~\cite{Jercher:2021bie,Jercher:2022mky}, mean-field theory applies generically as $\efd(k)\to\infty$ excedes the critical dimension in any case~\cite{Marchetti:2022igl,Marchetti:2022nrf}. This lays the theoretical foundations for applications of TGFT to phenomenology based on condensate states in a mean-field regime~\cite{Gielen:2013kla,Oriti:2015rwa}. \textcolor{black}{In particular, these results strongly lend support to the TGFT condensate cosmology program which has successfully extracted effective continuum cosmological dynamics from TGFT condensate states~\cite{Gielen:2013kla,Gielen:2013naa,deCesare:2016rsf,Pithis:2016wzf,Pithis:2016cxg,deCesare:2017ynn,Pithis:2019tvp,Oriti:2021rvm,Oriti:2021oux,Marchetti:2021gcv,Jercher:2021bie,Oriti:2023mgu}, see also Sections~\ref{sec:sakellariadou} and \ref{sec:marchetti} of this collection. As explained in the latter as well as in the introduction to this collection, this program is a particular instantiation of the \textit{hydrodynamics on superspace} framework.}

These results are based on Landau-Ginzburg theory as well as calculations of the functional renormalization group (FRG).
In the FRG one calculates flow equations for the effective average action in the scale $k$. 
To determine fixed points and the corresponding critical exponents, this equation can usually be mapped to an autonomous ordinary differential equation (ODE) by rescaling to dimensionless couplings $\tilde{\lambda}$, e.g. for local interactions of $n$ scalar fields $\lambda_n \propto k^{d - \frac{d-2}{2}n}\tilde{\lambda}_n$.
With non-local interactions in TGFT on  $r$ copies of the group $\mathbb{R}^\gd$ the same result applies but with $\efd =\gd(r-s)$ instead of $d$. Specifically we have found $s=1$ in the melonic regime~\cite{Pithis:2020sxm,Pithis:2020kio,Geloun:2023ray,Juliano:2024rgu} and $s=r/2$ is expected in the necklace regime on the grounds of~\cite{Carrozza:2017vkz}. 
However, if the group $G$ is compact with volume $V_G$, the FRG equation is a non-autonomous ODE in $k$ even upon rescaling.
It can however be brought to the standard, ``quasi-dimensionless'' form shifting any $k$-dependence into the effective dimension defined as
\begin{equation}\label{eq:effdim}
\efd(k) := \frac{ \partial \log F(k)}{\partial \log k}
\quad\textrm{ using a rescaling }\quad 
\lambda_n \propto k^n F(k)^{1-n} \tilde{\lambda}
\end{equation}
where $F(k)$ is a polynomial in $V_G\cdot k$ of order $\gd(r-s)$. 
Thus, for fixed $V_G$ the effective dimension~$\efd$ flows from its value on $\mathbb{R}^\gd$ at large~$k$ to zero at small~$k$.
In the FRG picture, mean-field theory is applicable if there is only a single relevant direction at the Gaussian fixed point which is true when $\efd > \crd = 4$. 
On a compact group, addition of $d_\phi>\crd$ matter degrees of freedom is necessary to fulfil this criterion.

In Landau-Ginzburg theory one can justify mean-field theory when Gaussian fluctuations around a constant vacuum field configuration $\Phi_0$ remain small, that is the ratio $Q$ of their average up to the correlation length scale $\xi\sim1/k$ with the average of $\Phi_0$ itself has $Q\ll1$ for $\xi\to \infty$.
For TGFT on non-compact groups we find for interactions of $n$ group fields with at most $r-s$ fold convolutions of their arguments 
\begin{equation}
Q \propto \bar{\lambda}^{\frac{2}{n-2}} 
    \mu^{-\frac{n}{n-2}}  V_\xi^{-(r-s)}
\end{equation}
where the $\bar{\lambda}$ is the coupling rescaled in an IR regularization parameter which is already removed in the equation.
On flat group domain, the scaling in the mass parameter $\mu\sim1/\xi^2$ gives the critical dimension $\crd = \frac{2n}{n-2}$ while the scaling in the domain volume integrated up to $\xi$, $V_\xi\sim\xi^{\gd}$, contributes the dimension $\efd=\gd(r-s)$ such that $Q \sim \xi^{\crd-\efd}$ and mean-field theory applies if $\efd>\crd$~\cite{Marchetti:2020xvf}.

The behaviour changes significantly with hyperbolic group geometry~\cite{Marchetti:2022igl}. 
For example, on $\text{SL}(2,\mathbb C) \cong \text{SU}(2) \times \text{A}^+ \times \text{SU}(2)$ the relevant contribution to the volume comes from the diagonal Cartan subgroup $\text{A}^+$ with measure $\sinh^2(\eta/a) \text{d}\eta/a$ including a curvature parameter $a$,
\begin{equation}
V_\xi = \frac{1}{4}\left(\sinh\left(\frac{2\xi}{a}\right) - \frac{2\xi}{a} \right)
\underset{\xi\to\infty}{\sim} \frac{1}{8} e^{\frac{2\xi}{a}}\,.
\end{equation}
As a consequence the $Q$ ratio is exponentially suppressed and mean-field theory is valid irrespective of $\gd$ and $r-s$.
Using the notion of scale-dependent effective dimension \eqref{eq:effdim}, one can again cast the result in the form $Q \sim \xi^{\crd-\efd(\xi)}$ but now with the function $F(k\sim1/\xi) \propto V_\xi$. 
This yields~\cite{Marchetti:2022nrf}
\begin{equation}
\efd(k\sim1/\xi) = (r-s)\frac{\cosh(\frac{2\xi}{a})-1}{\frac{a}{2\xi} \sinh(\frac{2\xi}{a})-1},
\end{equation}
which flows from $\efd(\xi) = 3(r-s)$ at small $\xi$ to $\efd(\xi)\to\infty$ at $\xi\to\infty$ (where the $3$ derives from the topological dimension of the $3$-hyperboloid $\mathrm{H}^3=\text{SL}(2,\mathbb C)/\text{SU}(2)$ being the hyperbolic part of $\text{SL}(2,\mathbb C)$). 
From this perspective, mean-field theory applies to TGFT on hyperbolic groups because the effective dimension outbalances any finite critical dimension towards large $\xi$, see also Refs.~\cite{Dekhil:2024ssa,Dekhil:2024djp} for further elaborations. These results provide strong support for the existence of a meaningful continuum gravitational regime in TGFT quantum gravity and the closely related spin foam models as well as lattice quantum gravity and crucially ties it to the presence of the Lorentz group. In particular, the obtained mean-field vacuum opens up the possibility to compare it with the Ashtekar-Lewandowski vacuum~\cite{Ashtekar:1993wf} and the Geiller-Dittrich vacuum~\cite{Dittrich:2014wpa} studied in LQG and spin foam gravity.

Moreover, the finding on the impact of the hyperbolic part of the Lorentz group allows for an educated guess of the phase structure even beyond Gaussian approximation. FRG calculations for local fields on $\mathrm{H}^3$ show that the flowing effective potential freezes around the curvature scale $a$~\cite{Benedetti:2014gja} such that the two phases of the mean-field regime describe the entire phase diagram.
While in TGFT on $\mathbb{R}^\gd$ higher orders in a derivative expansion may uncover further fixed points even for finite $\efd>\crd$ ~\cite{Pithis:2020kio}, we expect such structure to vanish when $\efd\to\infty$ on hyperbolic domains leaving us only with the mean-field continuum phase at large $\xi$.
While this is encouraging for IR phenomenology such as TGFT condensate cosmology~\cite{Gielen:2013kla,Gielen:2013naa,deCesare:2016rsf,Pithis:2016wzf,Pithis:2016cxg,deCesare:2017ynn,Pithis:2019tvp,Oriti:2021rvm,Oriti:2021oux,Marchetti:2021gcv,Jercher:2021bie,Oriti:2023mgu}, it still leaves ample room for interesting UV behaviour, e.g. new anisotropic UV fixed points~\cite{Juliano:2024rgu}, at small length scales, that is the actual quantum-gravity regime.

\subsection{Observables in Asymptotic Safety\\ \textit{by Renata Ferrero, Roberto Percacci and Antonio D.~Pereira}}\label{sec:ferrero}

The asymptotic safety program is an established continuum quantum-field-theory based  approach to quantum gravity that realizes the ultraviolet completion of quantum gravity via a non-trivial fixed point in the renormalization group (RG) flow. It replaces the infinitely many free parameters arising from the perturbative non-renormalizability of General Relativity by finitely many, thanks to the constraints imposed by the scale-invariance attained at the ultraviolet fixed point. The conjecture that gravity could be treated as an asymptotically safe quantum field theory was put forward by Weinberg in~\cite{Hawking:1979ig}. Yet it was thanks to the implementation of the Functional Renormalization Group (FRG)~\cite{Berges:2000ew,Delamotte:2007pf,Dupuis:2020fhh} to this program that it got momentum, see~\cite{Reuter:1996cp, Percacci:2017fkn, Reuter:2019byg}. FRG-based calculations have provided a compelling body of evidence for the existence of a suitable fixed point. Conversely, the synergy with other approaches such as Causal Dynamical Triangulations (CDT)~\cite{Loll:2019rdj} and tensor models~\cite{Gurau:2016cjo,GurauBook,Eichhorn:2018phj,Gurau:2024nzv} brings independent evidence for the existence of a suitable continuum limit in quantum gravity.

In a realistic scenario, both gravity and matter should be asymptotically safe. It is logically possible that matter quantum fluctuations could destroy the pure-gravity RG fixed point. Moreover, even the minimal coupling of matter to gravity induces matter-self-interactions as well as non-minimal interactions that should be asymptotically safe, see~\cite{Eichhorn:2012va,Dona:2013qba,Laporte:2021kyp,Eichhorn:2022gku}. In particular, those induced interactions cannot feature a free fixed point by symmetry reasons and therefore finding a suitable interacting fixed point for those interactions constitutes a fairly non-trivial test for the asymptotic safety scenario.

One issue with the application of the FRG to the AS program is that the couplings
that appear in the Lagrangian are not themselves observables and their dependence
on the cutoff is also not directly related to the momentum dependence of any observable~\cite{Donoghue:2019clr, Bonanno:2020bil}.

To study this in a simpler context, in~\cite{Buccio:2022egr}
a shift-invariant scalar field,
with 2- and 4-derivative kinetic terms and quartic derivative interactions
has been proposed.
These terms mimic the parts of the gravitational action that contain 1- and 2-curvatures.
It is found within the FRG that the theory flows from the 4-derivative free fixed point
to the 2-derivative free fixed point. AS appears as a limiting case of these
asymptotically free trajectories, when they approach a nontrivial fixed point.
These trajectories appear to be pathological, because the interaction energy
is negative, and there are ghosts but at least in the AS limit the ghost mass
goes to infinity.

To see how this picture relates to observables, the
2-to-2, one-loop scattering amplitude for the same system
has been calculated.
This can be compared to the one-loop approximation of the FRG, that can be easily
obtained by downgrading the full calculation, and leads essentially to the same
picture, except for numerical adjustments.
As generally expected, the dependence of the amplitudes on the external momenta
does not agree with the cutoff-dependence of the Lagrangian couplings,
except in two regimes.
At very low energy both become constants, and in the UV limit both reproduce the same
beta function. This proves the applicability of the FRG results in the high energy regime.
However, a deeper understanding of the physical behavior of this system requires
further investigation, in particular the calculation of the cross section.
This is because the standard LSZ formula cannot be straightforwardly applied
in the presence of 4-derivative propagators.

Another direction to build observables and incorporate them in the FRG approach to asymptotically safe gravity has been proposed in~\cite{Baldazzi:2021fye}.
It is known~\cite{Pagani:2016dof, Pagani:2016pad, Pagani:2017tdr, Pagani:2020ejb, Becker:2018quq, Becker:2019fhi}, that within the FRG the flow equation for composite operators can be applied to investigate scaling behavior of observables in quantum gravity. The composite operator formalism allows to calculate the expectation value of observables by solving a flow equation in which the flow parameter is the cutoff scale in the  IR $k$. The initial condition at the cutoff scale in the  UV $k = \Lambda$ is given by the expression of the observable as a function of the microscopic fields;  importantly, in asymptotic safety the UV cutoff can be taken to infinity  due to the existence of an interacting  fixed point. At the end point of the flow, when $k = 0$, the observable has evolved to the expectation value expressed as a function of the mean-field. We refer the reader to~\cite{Fehre:2021eob} for a more recent advancement in Lorentzian signature and a connection with the reduced phase space approach within the canonical formalism~\cite{Ferrero:2024rvi}.

In  gravitational theories, measurements are possible due to the presence of additional matter fields which provide a preferred frame of reference~\cite{DeWitt:1962kpz, Rovelli:1990pi, Rovelli:1990ph, Westman:2007yx, Goeller:2022rsx}\textcolor{black}{, see in particular the contributions in Section~\ref{sec:relationalism} on relationalism.} Indeed, the matter fields can be understood  as representing the readings on some physical coordinate scaffolding. All measurements of tensor or scalar fields referring to such physical coordinate systems are defined as ``relational observables''. 
Relational observables are therefore natural candidates for composite operators, which can be studied in AS. Moreover at the UV fixed point the corresponding composite operators will have universal scaling exponents. Physically these exponents should appear in the scaling behavior of correlation functions of relational observables at small distances less than the Planck length where effects of the fixed point scaling are expected. As such the computation of these exponents can serve as a way to compare with different approaches to quantum gravity. For example, the same exponents could be computed in other approaches to quantum gravity discussed in this collection, e.g. CDT, GFT or spin foam gravity.

Technically, they constructed a ``relational effective average action'' and derive its flow equation. In this way a natural criterion of choice of observables within an expansion of the flow equation arises: the flow equation and the fixed points automatically provide a subclass of consistent observables. They performed the first application to the flow of two relational observables corresponding to the inverse metric and the scalar curvature, where the physical coordinate system is composed of four massless scalar fields. As a result,  the scaling behavior of the relational observables at the UV fixed point is affected by  comparably small quantum corrections. This investigation was  also repeated in the minimal essential scheme~\cite{Baldazzi:2021ydj}, i.e., the scheme in which only the running of those couplings is considered, which affect the physical observables. 

This investigation could be extended in a number of ways. In a cosmological setting, for example, a dense dust of particles, which in the continuum limit becomes a continuous fluid, is commonly used to set up a physical coordinate system, and in this way gives a physical meaning to continuous tensor field, which can be measured in some region of spacetime.

\section{Relational physics in gravity and quantum gravity}\label{sec:relationalism}

A major challenge in all approaches to quantum gravity is the definition of observables, as already touched upon in the previous section. Gauge-invariant observables in gravity cannot be local from the perspective of the continuum spacetime manifold, necessitating a re-evaluation of how to assign gauge-invariant meanings to locality and events~\cite{Torre:1993fq}. One promising approach to the resolution of this issue, originally suggested by DeWitt~\cite{DeWitt:1962kpz}, is to define observables relationally. In this framework, relational observables are constructed as correlations between dynamical fields, where localization is determined via certain dynamical systems that function as physical reference frames~\cite{Giddings:2005id,Dittrich:2006ee,hoehn2018switch,hoehn2020equivalence,Gielen:2020abd}. Anticipating Section~\ref{sec:kirklin}, by introducing a field coordinatization of spacetime, one can show that, at the classical level, this enhanced (relational) notion of locality is consistent with key physical requirements, such as (relational) general covariance, and (relational) bulk microcausality.

A similar relational construction can be applied to quantum mechanical systems, where the degrees of freedom used as dynamical frames can exhibit quantum properties, such as entanglement or quantum superposition. Section~\ref{sec:giacomini} explores the physical implications of describing a system relative to these \emph{quantum} reference frames, see also~\cite{Giddings:2005id,Dittrich:2006ee,hoehn2018switch,hoehn2020equivalence,Gielen:2020abd}. While extending these techniques to quantum gravity remains an open question, notable progress has been achieved within the framework of reduced quantization, where a dynamical frame is selected prior to quantization. Section~\ref{sec:giesel} examines the implications of this reduced quantization approach in the context of Loop Quantum Gravity, with a particular focus on cosmological applications. 

\textcolor{black}{Note that while the contributions below can be viewed as standalone perspectives on their respective topics, they collectively contribute to the development of the relational strategy. In particular, they advocate for a notion of physical localization in field space rather than spacetime, which is an essential component of the hydrodynamics-on-superspace framework. In this way, the contributions also feed into its further development. Finally, note} that the concepts displayed in this section are of particular relevance to the discussions in Section~\ref{sec:phasetransitions} on phase transitions and emergent behavior in quantum gravity, as well as Section~\ref{sec:cosmology} on emergent cosmology \textcolor{black}{(see in particular Section~\ref{sec:marchetti})}. 

\subsection{Dynamical frames in gravity: observables, locality and covariance\\\textit{by Joshua Kirklin and Philipp Höhn}}\label{sec:kirklin}

The principle of general covariance is often colloquially stated as follows: \emph{all the laws of physics are the same in every reference frame}, where a `reference frame' is usually thought of as the system of local coordinates provided by some collection of experimental apparatus (such as rods and clocks) located in a region of spacetime. But a fundamental feature of any background-independent theory of gravity is that bulk diffeomorphisms, which move such apparatus around, and hence change the coordinates they provide, are gauge transformations. Thus, there is a tension between general covariance, bulk diffeomorphism invariance, and the physical significance of the notion of locality associated with reference frames. One way to resolve this tension is to give up on locality entirely, and resign oneself to a fundamental non-locality in gravity.

A more appealing alternative is to recognise the significance of \emph{dynamical} reference frames, which are those made up of physical degrees of freedom. This was already pointed out by Einstein in 1951, who wrote:
\begin{quote}
    ``The theory.... introduces two kinds of physical things, i.e., (1) measuring rods and clocks,
    (2) all other things, e.g., the electro-magnetic field, the material point, etc. This, in a
    certain sense, is inconsistent; strictly speaking measuring rods and clocks would have
    to be represented as solutions of the basic equations..., not, as it were, as theoretically
    self-sufficient entities....''
\end{quote}
Dynamical rods and clocks may be used to measure other degrees of freedom, by asking questions such as: when this rod measures a certain length, and that clock measures a certain time, what is the value of a certain component of a field? The so-called ``relational observables'' constructed in this way are completely gauge-invariant, and the version of locality provided by the dynamical coordinate systems of these physical rods and clocks consequently has a genuine physical significance. It is reasonable to hope that the tension noted above may be resolved by upgrading the usual locality associated with \emph{fixed} reference frames to the so-called \emph{relational locality} of \emph{dynamical} reference frames. Many approaches to exploring the consequences of this upgrade have been put forward in the literature (such as~\cite{Bergmann:1960wb,DeWitt:1962kpz,Rovelli:1990pi,Gambini:2000ht,Dittrich:2006ee,Giesel:2012rb,Giddings:2005id,Marolf:2015jha}, to mention just a few).

Relational observables are non-trivial observables associated with local events in spacetime. For relational locality to give a viable resolution to the tension, it is desirable to further establish the following:
\begin{itemize}
    \item \emph{Consistency with general covariance}: One should be able to say that all the laws of physics are the same in every \emph{dynamical} reference frame.
    \item \emph{Relational bulk microcausality}: Relationally local bulk observables should commute at spacelike separation.
    \item \emph{Gauge-invariant local subsystems}: There should be a physical notion of subsystems associated with relationally local subregions, suitable for information-theoretic and thermodynamic considerations. 
\end{itemize}
In~\cite{Goeller:2022rsx}, a detailed formalism establishing these properties in classical gravity, and unifying previous approaches, was proposed. The key ingredient introduced was a field-dependent coordinatization of spacetime. More precisely, this is a map $\mathscr{R}[\phi]:\mathscr{O}\to\mathcal{M}$, where $\phi$ denotes the configuration of the fields (including the metric), $\mathscr{O}$ is called the `local orientation space', and $\mathcal{M}$ is spacetime. The local orientation space should be thought of as the space of values that can be taken by the parameters of the reference frame. For example, a point in the bulk of spacetime can be reached by going a certain distance along a geodesic shot in a certain direction from a certain boundary point; the possible values for this distance, direction, and boundary point make up the local orientation space $\mathscr{O}$. The map $\mathscr{R}[\phi]$ takes us from an $o\in\mathscr{O}$ to the endpoint $x\in\mathcal{M}$ of the corresponding geodesic; by varying over $\mathscr{O}$ we trace out a region of spacetime~$\mathcal{M}$.

More generally, a map $\mathscr{R}[\phi]$ of this kind captures the fundamental features of the dynamical reference frames relevant in gravity. For relational locality to work, this map should be gauge-covariant, meaning if we act on the fields with a diffeomorphism, the coordinate system associated with the frame should move with the fields (this is obeyed for example by the boundary-anchored geodesic construction above). It can be concisely stated as: $\mathscr{R}[f_*\phi] = f(\mathscr{R}[\phi])$, where $f:\mathcal{M}\to\mathcal{M}$ is any bulk spacetime diffeomorphism.

As described in~\cite{Goeller:2022rsx}, gauge-covariant dynamical coordinate systems $\mathscr{R}[\phi]$ provide a powerful and general tool for understanding relational locality. Relationally local observables may be understood as tensor fields on the local orientation space $\mathscr{O}$, constructed as the pullback of covariant spacetime fields through $\mathscr{R}[\phi]$. Consistency with general covariance follows from the existence of dynamical `change of frames' maps $\mathscr{R}_{1\to 2}[\phi] = \mathscr{R}_2[\phi]^{-1}\circ \mathscr{R}_1[\phi]$ between two different coordinate systems $\mathscr{R}_{1,2}[\phi]:\mathscr{O}_{1,2}\to\mathcal{M}$ (defined on their spacetime overlaps, for injective coordinate systems), and through the construction of `relational atlases' and a `universal dressing space'. This change of frames map is itself gauge-invariant, and hence so is this updated version of general covariance. Relational microcausality may be established in certain cases by using Peierls' bracket (although some challenges with boundary conditions remain for reference frames that transform non-trivially under large diffeomorphisms -- see~\cite{Fighting} for an analogous discussion in QED). In addition, the phase spaces of relationally local subsystems may be defined in terms of subregions $\mathscr{U}\subset\mathscr{O}$ of the local orientation space, rather than of spacetime.

This formalism is classical; to approach the same questions in \emph{quantum} gravity, the classical reference frames must be converted into \emph{quantum} reference frames, in order to account for the quantum fluctuations of spacetime. Finding the best way to do so remains a challenging open area of research but recent steps in this direction, informed by Von Neumann algebraic constructions, have been made (see~\cite{DeVuyst:2024pop}). These involve very simplified reference frames, and although a quantum construction at the level of generality and power in~\cite{Goeller:2022rsx} has yet to be put forward, it seems that a fruitful path forward is revealing itself.

\subsection{Quantum Reference Frames:\\ A Relational Perspective on Non-classical Spacetime\\\textit{by Flaminia Giacomini}}\label{sec:giacomini}

The two fundamental theories of modern physics, Quantum Theory (QT) and General Relativity (GR), rest on the classical notion of spacetime. In Quantum Gravity (QG), in which both quantum and gravitational effects are relevant, it is widely expected that the very notions of time, space, and causality will be profoundly different to the ones we currently use in physics, and hence that our physical intuition based on spacetime should drastically change. 

Background independence is often mentioned among the desirable features that a theory of QG should have and is connected to relationalism~\cite{butterfield2001spacetime, anderson2017problem}. A relational theory that also has quantum features should allow for more general (i.e.\,quantum) relations between the different perspectives. For instance, it should extend the notion of  diffeomorphism invariance to allow for some sort of quantum coordinate systems, namely coordinate systems which are in a quantum relationship relative to each other. Achieving such a quantum formulation of relational perspectives is extremely hard in a full gravitational framework. This is why, usually, QG approaches first formulate such a relational picture classically and then quantize it, see for instance~\cite{dewitt1967quantum, Kuchar:1990vy, Brown:1994py, Brown:1995fj, rovelli_quantum, Rovelli:2004tv, rovelli_relational, Dittrich:2004cb, Tambornino:2011vg, hoehn2018switch, hoehn2019switching, hoehn2019trinity, hoehn2020equivalence, Goeller:2022rsx}.

A complementary perspective is provided by research in quantum foundations, where work on the notion of Quantum Reference Frames (QRFs)~\cite{Giacomini:2017zju, perspective1, perspective2, giacomini2019relativistic, de2020quantum, krumm2020, ballesteros2020group, streiter2020relativistic,  mikusch2021transformation, castro2021relative, de2021perspective, de2021entanglement, hoehn2021quantum, hoehn2021internal,  giacomini2022udw} has started to incorporate such a relational view in quantum mechanics and is now generalising it to include additional gravitational elements~\cite{castro2020quantum, giacomini2020einstein, giacomini2021spacetime, giacomini2021quantum, cepollaro2021quantum, de2021falling, Christodoulou:2022knr, Kabel:2022cje, Overstreet:2022zgq}. The general idea of QRFs is to associate a reference frame to a quantum system, which can be in a quantum superposition or entangled from the perspective of a different quantum system. In this sense, QRFs can be viewed as an instance of relational quantum space (in some cases, spacetime). Here, we mainly refer to a specific literature on QRFs, e.g. Refs.~\cite{Giacomini:2017zju, perspective1, perspective2, giacomini2019relativistic, de2020quantum, krumm2020, ballesteros2020group, streiter2020relativistic,  mikusch2021transformation, castro2021relative, de2021perspective, de2021entanglement, hoehn2021quantum, hoehn2021internal,  giacomini2022udw, castro2020quantum, giacomini2020einstein, giacomini2021spacetime, giacomini2021quantum, cepollaro2021quantum, de2021falling, Kabel:2022cje, Overstreet:2022zgq, delaHamette:2022cka} but for completeness see also Refs.~\cite{aharonov1, aharonov3, brs_review, spekkens_resource, palmer_changing, smith_quantumrf, poulin_dynamics, busch_relational_1, busch_relational_2, busch_relational_3, angelo_1, Hardy:2018kbp, zych2018relativity, hardy2020implementation}.

QRFs allow one to generalize a reference frame transformation to a \emph{quantum superposition of reference frame transformations}. This formulation of QRFs has several features which could be useful in a quantum-gravitational setting, and specifically to study physics without a classical spacetime structure: 
\begin{enumerate}
	\item In standard quantum mechanics with Galilean symmetries, it was shown~\cite{Giacomini:2017zju} that a quantum product state can be mapped to an entangled state via a QRF transformation and viceversa. More in general, entanglement and superposition depend on the choice of the QRF. This relative localization property also applies to the time localization of events, as measured by quantum clocks in a classical gravitational field or interacting gravitationally~\cite{castro2020quantum, giacomini2021spacetime, cepollaro2021quantum}. In addition, the state of the gravitational field can be mapped, under certain conditions, from a quantum superposition state to a well-defined state (namely, one corresponding to a classical configuration of gravity) and viceversa~\cite{de2021falling, Overstreet:2022zgq};
	\item Fundamental principles of physics, such as the covariance of physical laws~\cite{Giacomini:2017zju} and the Equivalence Principle~\cite{giacomini2020einstein, giacomini2021quantum, cepollaro2021quantum}, can be shown to be valid in a more general sense under certain QRF transformations and for a superposition of geometries. This feature points towards the possibility of formulating a quantum diffeomorphism structure~\cite{Goeller:2022rsx, delaHamette:2022cka}, and of considering different or more general symmetry groups~\cite{ballesteros2020group} than those in standard physical theories;
	\item QRFs give physical meaning to relational quantities in terms of operations on physical systems~\cite{giacomini2019relativistic, streiter2020relativistic, giacomini2022udw}. This property is particularly useful in contexts in which a relevant observable does not have a clear or unambiguous physical meaning. In relativistic quantum theory, for instance, this was the case for the covariant spin operator, for which no operational procedure existed to identify it (namely, a relativistic Stern-Gerlach experiment). In the rest frame of any physical system, however, the operational identification of spin is unambiguous. With a QRF transformation it is possible to map the spin operator from the rest frame of a quantum system in a superposition of velocities to the laboratory frame, and hence provide the required operational identification~\cite{giacomini2019relativistic}. This feature could be useful in QG scenarios to operationally relate observables to quantities with a direct physical interpretation.
\end{enumerate}

In conclusion, this research identifies a bottom-up approach which allows us to combine elements of QT and GR and test their internal consistency in concrete physical scenarios. Via this procedure, it provides us with fundamental principles which should be retained or modified at the interface between these two theories. It is an open question how to combine this approach with QG techniques but, if successful, this could provide indications on the fundamental structure and observable consequences that a QG theory should have.

\subsection{Dynamical Reference Frames in Canonical Quantum Gravity\\\textit{by Kristina Giesel}}\label{sec:giesel}

The choice of dynamical, and hence physical, reference frames plays an important role in gravitational systems, as it allows the definition of a physical notion of spatial and temporal coordinates. In  general relativity, the relational formalism~\cite{Rovelli:1989jn,Rovelli:1990ph,Rovelli:2001bz,Vytheeswaran:1994np,Dittrich:2004cb,Thiemann:2004wk,Dittrich:2005kc,Pons:2009cz,Pons:2010ad} can be used to implement a particular choice of dynamical reference frames, construct the corresponding Dirac observables, and derive the associated physical phase space in the canonical theory. In models of quantum gravity, there is the additional freedom to choose dynamical reference frames at either the classical or quantum level, and thus to work with classical or quantum reference frames. In the canonical setup, this choice corresponds to either a Dirac or a reduced quantization for the model under consideration. With a Dirac quantization, one can follow the perspective-neutral approach discussed in~\cite{delaHamette:2021oex} for simpler systems than general relativity.

Since this approach is very difficult for full quantum gravity models, as it requires knowledge of the physical Hilbert space, the focus here is on the reduced phase space quantization approach. In this case, one introduces some dynamical reference systems at the classical level and reduces the constraints there. Then one quantizes only the reduced physical phase space by finding a representation of the corresponding algebra of Dirac observables. This has been done in~\cite{Giesel:2007wn,Domagala:2010bm,Husain:2011tk,Husain:2011tm,Giesel:2012rb,Giesel:2016gxq,Ali:2018vmt,Han:2019vpw} for various dynamical matter reference frames based on the seminal work in~\cite{Brown:1994py,Kuchar:1990vy,Kuchar:1995xn} and in the linearised gravity framework for geometric clocks, these are dynamical reference frames consisting only of gravitational degrees of freedom, see for instance in~\cite{Fahn:2022zql}. In these models, one thus chooses a classical reference frame and quantizes an observer-dependent quantum field theory of the physical sector. In general, geometric clocks are more difficult to handle because their underlying algebra of observables is more complicated than that of suitable matter reference frames. In this context, the question naturally arises as to what physical properties such observer-dependent models have and how the choice of a particular observer and thus a dynamical reference frame might influence these properties. A first step towards analyzing such questions was taken in~\cite{Giesel:2020raf}, where different choices of clocks were investigated in the context of a reduced phase space quantization in loop quantum cosmology, see also~\cite{Gielen:2021igw} for work on the choice of clocks in the context of the singularity resolution in quantum cosmology. All the models considered in~\cite{Giesel:2020raf} involve gravity and an inflaton and then a dynamical clock chosen as either a dust or scalar field. The reduced quantization was carried out in the framework of loop quantum cosmology and the corresponding Schrödinger-like equations encoding the dynamics were obtained.
Note that the models that are compared in~\cite{Giesel:2020raf} are physically distinguishable because they involve a coupling of different matter to gravity and the inflaton similar to what has be done in~\cite{Giesel:2007wn,Domagala:2010bm,Husain:2011tk,Husain:2021ojz,Giesel:2012rb,Giesel:2016gxq} for full general relativity.  If we choose a fixed model and within that model we switch the clock, i.e. choose a subset or another subset of degrees of freedom as the dynamical reference frame to compare choices of clocks, then this corresponds to the analysis done for instance in~\cite{Gielen:2021igw} in quantum cosmology.

To compare the physical properties of these models, a Starobinsky potential was chosen and the corresponding effective models were considered. As the work in~\cite{Giesel:2020raf} shows, both the scalar field and the dust clocks serve as good dynamical reference frames, since there are initial conditions for which the influence of the clocks is tiny, these are initial conditions where the energy density of the clocks is much smaller than the one of the inflaton. For example, to test the influence, the number of e-foldings was calculated and compared to the result of the APS model~\cite{Ashtekar:2006rx} for comparable initial conditions set at the bounce, where a value of the volume of $10^3$ in Planck units was chosen. In the APS model one obtains $63.9$ e-foldings~\cite{Li:2019ipm}, while for the model that additionally includes the dust clock  one gets  $63.1$ and for the scalar field clock one obtains $63.8$ e-foldings. Note that the APS model can be understood as one in which the inflaton itself is chosen as the dynamical clock. It is also shown that the number of e-foldings, as expected, decreases when the energy density of the clock is increased in a range from  $10^{-8}$ to $10^{-4}$. For the same set of initial conditions this decrease is less significant for the scalar field than for the dust clock, and for the latter there is even an upper limit to its energy density around $1.4 \times 10^{-4}$, for the choice of initial conditions used in~\cite{Giesel:2020raf}, where there is no inflation any longer. This is an example that shows that although we choose a dynamical classical reference frame, it has an impact on the final quantum cosmology model because it is dynamically coupled with the degrees of freedom that are quantized. 

Since a given choice of a dynamical reference frame can be associated with a given gauge fixing, where the associated Dirac observables can be understood as gauge invariant extensions of those phase space variables for which observables are constructed, one possibility to move towards quantum reference frames in gravitational systems by implementing gauge fixings at the quantum level. Again, as a first step in~\cite{Giesel:2021rky}, the construction of effective models for polymerised gravitational systems was used. Since gauge fixing and polymerization are generally expected to be non-commutative, the work in~\cite{Giesel:2021rky}  analyzed in which special cases such commutativity exists and, if not, how a consistent gauge fixing can be performed in polymerised models. The latter becomes particularly relevant when the lapse function and/or the shift vector depend on phase space variables that are polymerised, as is the case for some models in the literature, see for instance~\cite{Corichi:2015xia,Kelly:2020uwj,Husain:2021ojz,Gambini:2020nsf}. The work in~\cite{Giesel:2021rky} discusses a few lemmas for a class of dynamical reference models and states under which assumptions gauge fixing and quantization commute. It turns out that only in very special situations the two are commuting, and such situations occur more naturally for matter than geometrical clocks as the latter are usually polymerised. One of the conclusions one can draw from the results in~\cite{Giesel:2021rky} is that if at least one of the gauge fixing conditions depends on the temporal coordinate and involves polymerised degrees of freedom then gauge fixing and polymerization will not commute. 
Interestingly, all dynamical matter reference models in~\cite{Giesel:2007wn,Domagala:2010bm,Husain:2011tk,Husain:2011tm,Giesel:2012rb,Giesel:2016gxq} fall into a class of models, for which gauge fixing and quantization commute and thus classifying this type of matter reference frame as a special group of dynamical reference frames. For other models, the work in~\cite{Giesel:2021rky} shows that if commutativity does not exist, some caution is needed when quantizing lapse and shift, since in general these polymerised counterparts cannot be given simply by polymerising the classical lapse and shift. Rather, the form of the polymerised lapse and shift in a given model can be determined from the stability of the (polymerised) gauge fixing condition under the effective dynamics.  This strategy is applied to several existing models in the literature~\cite{Corichi:2015xia,Kelly:2020uwj,Husain:2021ojz} and compared with results from the literature. Furthermore, it is discussed how to reverse engineer a given (polymerised) gauge fixing when working with a given polymerised lapse function and shift vector, and what problems may arise in this context, e.g. that the reverse engineered gauge fixing may not merge into the classical gauge fixing we started with when the limit of a vanishing polymerization parameter is considered.

The analyses carried out in~\cite{Giesel:2020raf,Giesel:2021rky} are, of course, only first steps but they already show that the understanding of the physical properties of quantum gravity models is not independent of whether and how the choice of a dynamical reference frame, be it classical or quantum mechanical, influences the model. Moreover, in the models discussed here, boundaries have not played an important role. An interesting question is therefore also what are suitable dynamical reference frames for gravitational systems with boundaries, see~\cite{Carrozza:2021gju,Kabel:2023jve} for work in this direction, and how do these reference frames relate to the matter and/or geometrical dynamical reference frames chosen when boundaries are absent. This opens up interesting directions for future research aimed at gaining a deeper understanding and more insight into the relationship between classical and quantum reference frames and their properties, in general.

\section{Emergent cosmology from quantum gravity}\label{sec:cosmology}

There are a number of reasons suggesting that a complete physical understanding of our Universe requires a quantum description of gravity. 
First, under very general conditions, GR predicts the presence of an initial singularity~\cite{Calcagni2017,Calcagni:2022ssd}, which is expected to take place at energy scales where QG effects are dominant.
Second, in the widely accepted inflationary scenario, deviations from homogeneity and isotropy are believed to be seeded by primordial quantum fluctuations~\cite{Baumann:2009ds}. Therefore, a detection of primordial tensor perturbations would confirm the quantum nature of the gravitational field~\cite{maggiorebook}.
On top of that, the typical scale of perturbations that we observe today turns out to be effectively trans-Planckian when traced back to the onset of inflation, and thus might have been directly affected by QG physics~\cite{Martin:2000xs}.
Finally, QG might help us gain some physical insights on the nature of dark matter and dark energy~\cite{Padmanabhan:2014nca}, and potentially help resolving~\cite{Perez:2020cwa} cosmological tensions~\cite{Abdalla:2022yfr}. The interplay between QG, inflation and dark energy is discussed in more detail in Sections~\ref{sec:sakellariadou} and~\ref{sec:brandenberger} below.

On the other hand, cosmological applications are particularly appealing for QG. From the experimental side, the ever-increasing accuracy of cosmological observations can provide precision tests of gravitational theories~\cite{Planck:2018nkj}. From the theoretical side, cosmological geometries are particularly simple, since classically they are characterized by a high degree of symmetry due to an averaging of the local properties of the Universe over very large scales.

Despite the simple mathematical description of the Universe on large scales, it is still remarkably challenging to extract cosmology from full QG. The main reason for this is that cosmology deals with typical scales which are much larger than those at which the microscopic QG degrees of freedom (whatever they may be) operate. Therefore, one should expect cosmological physics to be only a macroscopic phenomenon, emerging from the collective behavior of the fundamental QG quanta. Physically understanding this emergence process is at the core of the problem of extracting cosmology from QG. When doing so, however, one is forced to face  deeply intertwined technical and conceptual issues. 

One might expect such an emergence process to be associated to some form of coarse-graining, and thus to be described via \textcolor{black}{mean-field or} (non-perturbative) renormalization group analyses of the underlying QG theory~\cite{Oriti:2013jga,Carrozza:2016vsq,Oriti:2021oux,Marchetti:2022nrf,Eichhorn:2018phj,Benedetti:2022ots} (see also Section~\ref{sec:phasetransitions}) but performing such an analysis on a typical QG model is incredibly complicated (it is in fact, already highly non-trivial for most standard QFTs on flat spacetime). On top of that, any physical interpretation of the coarse-graining procedure is tightly related to spacetime notions. An important example is the renormalization group scale according to which one might want to define the cosmological coarse-graining, and in particular, with respect to which one would describe the average homogeneity of the Universe, see for instance~\cite{Carrozza:2016vsq,Marchetti:2020xvf,Pereira:2019dbn} and~\cite{Eichhorn:2021vid} for further discussions.

From a perspective where spacetime and geometry are emergent, associated notions such as (spacetime) covariance and unitarity become non-trivial. This is explored through two explicit examples in Section~\ref{sec:bojowald}. Similarly, crucial cosmological concepts like homogeneity, isotropy, and cosmic expansion—foundational to our theoretical understanding of the Universe—should also be viewed as emergent. Consequently, these concepts offer only limited insights into the process of cosmic emergence itself.
Still, even though these challenges—rooted in the intrinsically pre-geometric nature of the QG degrees of freedom—complicate a complete mathematical modeling of the cosmic emergence process, they provide useful guidance for developing physically reasonable approximations. These approximations can help us navigate technical difficulties, leading to a mathematically controlled framework. Such a framework, though approximate, may yield valuable insights into the cosmic emergence process and iteratively refine our understanding and approximation methods.


For such a program to be successful, however, we should be careful to formulate the principles on which our approximations are based in a language appropriate to the fundamental QG theory. Going back to the example of homogeneity, isotropy and cosmic expansion, they should not be formulated in terms of dependence (or independence) with respect to some time (or space) coordinate, as these quantities have completely disappeared in the fundamental QG theory in virtue of background independence. Rather, they should be formulated in relational terms, as dependence (or independence) of physical quantities on various components of a physical reference frame~\cite{Rovelli:1990ph,Dittrich:2004cb,Dittrich:2005kc,Tambornino:2011vg,Goeller:2022rsx}, see also Section~\ref{sec:relationalism} for further details on the relationalist perspective.

\textcolor{black}{All of the above arguments apply in particular to a class of QG models called Group Field Theories (GFTs)~
\cite{Oriti:2009wn,Oriti:2011jm,Krajewski:2011zzu,Freidel:2005qe}. The process of cosmic emergence within this approach} \textcolor{black}{naturally fits into the \textit{hydrodynamics on superspace} framework, and} \textcolor{black}{is reviewed and discussed in Section~\ref{sec:marchetti}.} \textcolor{black}{ Some key related results are also highlighted in Section~\ref{sec:sakellariadou}.} \textcolor{black}{These findings are directly connected to the contribution on condensate states in GFT presented in Section~\ref{sec:thuerigen} of this collection.}

\textcolor{black}{Potentially, the arguments outlined in this introduction} \textcolor{black}{also apply to a non-perturbative definition of superstring theory, as described by the BFSS matrix model~\cite{Banks:1996vh} which is discussed in a cosmological context in Section~\ref{sec:brandenberger}. It is left to future research to investigate how the latter section could fit in detail into this novel perspective on cosmology from quantum gravity.
}

\subsection{On the affinities between Cosmology and Quantum Gravity \\\textit{by Mairi Sakellariadou}}\label{sec:sakellariadou}

There is a strong interplay between quantum gravity and cosmology. On the one hand, quantum gravity offers a framework to build a cosmological model based on a fundamental theory; on the other hand, cosmology is the ideal test bed for a quantum theory of spacetime geometry. Despite its enormous success, cosmology still faces some fundamental questions and quantum gravity may be the road to obtain some answers. In spite of several approaches (perturbative as well as non-perturbative ones) a successful quantum theory of spacetime, capable of explaining our universe, is still lacking. The interplay between these two fields of very active research is therefore very strong.

A successful cosmological model should satisfactorily address the emergence of space and time, the origin of an early-time exponential expansion, the origin of a late-time accelerated expansion, and the origin of cosmological perturbations. Certainly, spacetime dimensionality is another issue begging for an explanation. The standard cosmological scenario is the so-called $\Lambda$CDM cosmological model, a phenomenological model based on General Relativity, supplemented by two unknown ingredients (cold dark matter and dark energy), and complemented by an inflationary scenario. However, despite many efforts we still do not know what the dark matter is, the most probable candidates being primordial black holes and axions, none of which has yet been confirmed by observational data. Regarding inflation, while its dynamics are fixed, its origin is unknown and particular initial conditions may be required for the onset of inflation~\cite{Calzetta:1992gv,Calzetta:1992bp,Germani:2007rt}. Indeed, despite numerous studies and a variety of (basically ad hoc) inflationary models, inflation still remains a paradigm in search of a theory. Finally, the nature of dark energy is another big unknown~\cite{Brax:2017idh}. 

Quantum gravity corrections are believed to cure the issue of the initial singularity that plagues any classical theory of gravity~\cite{Hawking:1970zqf}. Indeed, several quantum gravity proposals can achieve resolution of the initial singularity. One may for instance see this achievement in the context of string theory (e.g., string gas scenario~\cite{Brandenberger:1988aj} (see also Section~\ref{sec:brandenberger}) or pre-big-bang cosmology~\cite{Gasperini:1996fu}) or within a non-perturbative approach to quantum gravity (e.g., loop quantum cosmology~\cite{bojowald2001absence} or group field theory~\cite{deCesare:2016axk,deCesare:2017ztf}, see also Sections~\ref{sec:thuerigen} and~\ref{sec:marchetti} of this collection).

Whilst several quantum gravity proposals can achieve resolution of the initial singularity, the onset of inflation has been in general seen by  assuming the existence of a scalar field and studying its dynamics. However, an early era of accelerated expansion should be drawn from quantum gravity corrections, in particular since inflation is supposed to have taken place during the very early stages of the universe when quantum corrections should not be neglected. Hence the ad hoc consideration of an inflaton field within a quantum gravity framework may not be the proper way to investigate the early exponentially expansion of the universe.

Indeed, quantum gravity may provide a mechanism (through quantum gravity effects) to explain the early exponential expansion of the universe. 
For instance, if one considers the emergence of continuum spacetime from discrete (pre-geometric) quantum structures, based on the idea of quantum space as a condensed matter system~\cite{oriti2007group}, one can show that the universe undergoes a cyclic (and non-singular) evolution, and its volume has a positive minimum, corresponding to a bounce~\cite{deCesare:2016axk}. In addition, to achieve an exponential expansion in the absence of an inflaton field with fine-tuned potential, one has to consider within this framework, a particular type of interaction between the quanta of geometry~\cite{deCesare:2016rsf,Pithis:2016cxg}, see also~\cite{Pithis:2019tvp}. Thus, observational data guide us in constructing the quantum gravity model.

Approaches to quantum gravity have been also providing an explanation for the spacetime dimensionality. This can be seen for instance in the framework of strings and branes~\cite{Brandenberger:1988aj,Sakellariadou:1995vk,Durrer:1998sv,Nelson:2008eu,Nelson:2008sv}.

There is a wide variety of very precise astrophysical and cosmological data, against which we can (and should) test predictions of quantum gravity proposals, since such proposals must not only be mathematically consistent but they must also model/explain the universe in which we live. In doing so however one should keep in mind that often (almost always) we do not test the full theory but the cosmological model (namely after assuming several symmetries) inspired by the full quantum theory. Similarly, we should be aware that in principle any raw observational data are explored within a particular framework, and hence the tests are just consistency tests of the given framework. Hence, one has to explore data within different frameworks, which is often a rather tedious enterprise.
Nevertheless, we live within a huge laboratory, our universe, which offers us precise tests to investigate whether a mathematically consistent quantum gravity theory can explain the universe in which we live.
In particular, gravitational waves research~\cite{GravWaveResearch} is a novel and powerful way to test not only the validity of a classical theory of gravity but also of quantum gravity proposals (e.g,.~\cite{Calcagni:2019ngc,Calcagni:2019kzo}).

\subsection{Two examples of emergent time\\\textit{by Martin Bojowald}}
\label{sec:bojowald}
The general appearance of time in physical theories implies several strong
consistency conditions that are easily satisfied by a classical number, such as
a coordinate but are much harder to realize if time is emergent from a
fundamental quantum object. The main conditions are implied by mathematical
relationships of time with space in a generally covariant setting (or any
spacetime description), and by unitarity requirements on time evolution in the usual
probabilistic formulation of quantum mechanics. Both types of conditions are
illustrated by the two examples in this section.

Time is more than a simple parameter that could appear, for instance, from
deparameterization of a Hamiltonian constraint in a canonical theory or from a
hydrodynamical treatment, and spacetime is more than a collection of
parameters that could appear from deparameterization by multiple fields. An
important geometrical feature that characterizes time and governs its
relationships with space is given by the signature of spacetime. In a
canonical theory, this property is determined by a specific sign choice in the
commutators of hypersurface deformations in a spacetime foliation. In
particular, two deformations along the normal direction of a spacelike
hypersurface with two different position-dependent displacements, $N_1$ and
$N_1$, have to commute according to
\begin{equation} \label{TT}
  [T(N_1),T(N_2)] = S(\beta q^{ab}(N_1\partial_bN_2-N_2\partial_bN_1)),
\end{equation}
where $T$ is a normal deformation and $S$ a tangential deformation, using the
spatial metric $q_{ab}$ on a spatial slice. The phase-space function $\beta$
is a constant sign in a classical theory, depending on the signature.

Quantum effects modify $T$ because it contains the Hamiltonian. Covariance in
a consistent spacetime theory in which time has a chance to emerge then
requires that a commutator of the form of Eq.~\eqref{TT} must survive an inclusion of
such effects. Such a modified theory is first-class because the constraints
corresponding to $T$ and $S$ commute on-shell but additional conditions are
required for covariance~\cite{Bojowald:2018xxu}. If $\beta$ is no longer constant, the theory is consistent with a covariant spacetime line element only if $|\beta|q_{ab}$ is used as the spatial metric, and the
signature must be adjusted according to ${\rm sgn}(\beta)$. Non-trivial,
highly restrictive conditions on $|\beta|q_{ab}$ ensure that it can play the role of a spatial metric.

Unitarity is another consistency condition that has, for a long time, stood in the way of successful constructions of fully dynamical, self-interacting time models. The usual choices of deparameterization require a special matter field with a simple Hamiltonian that is linear in the momenta (and therefore non-relativistic) or free of self-interactions or even mass. However, it is possible to construct a consistent evolution based on self-interacting time
models, for instance with a Hamiltonian $p_{\phi}^2 +\lambda\phi^2$ for a
clock variable $\phi$. The construction shows that a consistent notion of time
$\tau$ may emerge from interactions with such a clock but the fundamental and
possibly oscillating clock field $\phi$ is not the same as the emergent
monotonic time $\tau$. New physical effects include additional decoherence
implied by a fundamental clock, which gets weaker for a faster clock with
large $\lambda$. A comparison with precision measurements of time provides an
upper bound of
$T_{\rm C}< 2\cdot 10^{-33}\,{\rm s} \approx 0.5 \cdot 10^{11} t_{\rm P}$ on
the fundamental clock period~\cite{Wendel:2020hqv,Bojowald:2021uqo}.

\subsection{Emergent Cosmology from (T)GFT Condensates\\\textit{by Luca Marchetti}}\label{sec:marchetti}

\textcolor{black}{The following contribution gives a lightning review of the TGFT condensate cosmology approach. This will clarify how the topic of emergence of continuum geometry is addressed therein, see also Section~\ref{sec:thuerigen}, and how the relationalist perspective, discussed in Section~\ref{sec:relationalism}, is exploited to provide an approximate description thereof together with a deparametrized account of its dynamics in a cosmological context. In this way, it explains how this approach corresponds to a specific realization of the \textit{hydrodynamics on superspace} framework.}

To start off, consider that in $4$ dimensions the fundamental quanta of GFTs are quantum tetrahedra decorated with discretized geometric and matter field data. Being GFTs proper QFTs, one can construct a Fock space out of the single-tetrahedron Hilbert spaces~\cite{Oriti:2013aqa,Gielen:2013naa,Jercher:2021bie,Sahlmann:2023plc}. The field operator $\hat{\varphi}$ destroying one GFT quantum is then defined on the following domain:
\begin{equation*}
    \mathcal{D}=\text{superspace}/\text{geometricity constraints}\,,
\end{equation*}
where with superspace we mean the space of discretized (matter and geometry) fields, and the geometricity (i.e.\ simplicity and closure) constraints are imposed in order to guarantee the interpretation of the GFT quanta as \qmarks{atoms} of a $3$-dimensional space. Typically, $\mathcal{D}=M\times G^4/\text{geometricity constraints}$, where $M$ is the matter part of the domain (for instance, $M=\mathbb{R}^n$ for $n$ scalar fields) and $G=\mathrm{SL}(2,\mathbb{C})$ represents the local gauge group of gravity\footnote{Depending on the details of the imposition of the geometricity constraints, one could also identify $G$ as an $\mathrm{SU}(2)$ subgroup of $\mathrm{SL}(2,\mathbb{C})$.}.

\paragraph{Cosmology as (T)GFT hydrodynamics.}
For the reasons mentioned above, performing a non-perturbative renormalization group analysis of GFT models, and thus computing the quantum effective action is currently out of reach. One must therefore look for approximate methods, the simplest one of which being the \emph{mean-field approximation}\footnote{\textcolor{black}{We refer the reader to Section~\ref{sec:thuerigen} for a discussion on the validity of the mean-field approximation. 
}}~\cite{Pithis:2018eaq,Pithis:2019mlv,Marchetti:2020xvf,Oriti:2021oux,Marchetti:2022igl,Marchetti:2022nrf,Dekhil:2024ssa,Dekhil:2024djp}. It consists of a saddle point evaluation of the partition function of the system and in the consequent approximation of the effective action via the classical GFT action. From the perspective of the quantum many-body system, this is equivalent to imposing only averaged quantum \qmarks{equations of motion} on \emph{condensate states}~\cite{Gielen:2013naa, Oriti:2016qtz,Gielen:2016dss,Marchetti:2020umh,Jercher:2021bie}, that is
\begin{equation}\label{eqn:meanfield}
    \left\langle\frac{\delta S[\hat{\varphi},\hat{\varphi}^\dagger]}{\delta \hat{\varphi}(\mathcal{D})}\right\rangle_{\sigma}=0\,,\qquad \ket{\sigma}=\mathcal{N}\exp\left[\int\diff \mathcal{D}\,\sigma(\mathcal{D})\hat{\varphi}^\dagger(\mathcal{D})\right]\ket{0}\,,
\end{equation}
where $\left\langle\cdot\right\rangle_\sigma$ represents an average on the state $\ket{\sigma}$, $\diff\mathcal{D}$ is an appropriate measure on the domain $\mathcal{D}$ defined above and $\mathcal{N}$ a normalization constant. The above mean-field equations are the QG counterpart of the Gross-Pitaevskii (GP) equation for quantum fluids~\cite{Oriti:2016qtz}. As in the theory of quantum fluids, these equations set the stage for a hydrodynamic description of the system~\cite{pitaevskii2003bose}. 

A crucial role in this hydrodynamic description is played by the condensate wavefunction $\sigma(\mathcal{D})$, determined by solving the first equation in expression \eqref{eqn:meanfield}. Indeed, expectation values of collective observables (represented for instance as $1$-body operators on the GFT Fock space)~\cite{Oriti:2016qtz,Marchetti:2020umh}
\begin{equation}\label{eqn:hydrovariables}
    \langle \hat{O}\rangle_\sigma=\int\diff\mathcal{D}\diff\mathcal{D}'\,O(\mathcal{D},\mathcal{D}')\left\langle\hat{\varphi}^\dagger(\mathcal{D})\hat{\varphi}(\mathcal{D}')\right\rangle_\sigma=\int\diff\mathcal{D}\diff\mathcal{D}'\,O(\mathcal{D},\mathcal{D}')\bar{\sigma}(\mathcal{D})\sigma(\mathcal{D}')
\end{equation}
take the form of hydrodynamic variables, i.e.\ averages of the microscopic data $O(\mathcal{D},\mathcal{D}')$ weighted by $\sigma$. From this perspective, thus, $\sigma$ assumes the role of a distribution function over $\mathcal{D}$. Therefore, it determines the hydrodynamic and macroscopic properties of the system and encodes the collective behavior of the fundamental QG quanta. 

It is important to remark, here, that it is at the level of the hydrodynamic variables (and thus of the condensate wavefunction) that one can implement emergent notions as (relational) homogeneity, isotropy, and time evolution. Still, notice that the hydrodynamic variables \eqref{eqn:hydrovariables} are in general not relationally localized~\cite{Marchetti:2020umh,lucaed}. 

To introduce a notion of relational \qmarks{time} localization with respect to a desired physical clock (for instance, a minimally coupled, massless and free scalar field $\chi^0$ from the perspective of the corresponding classical theory), one could consider states whose condensate wavefunction is sharply peaked on a certain (arbitrary) value $x^0\in\mathbb{R}$ of $\chi^0$~\cite{Marchetti:2020umh}. For instance, this can be achieved by assuming that $\sigma$ factorizes into a peaking function $\eta$ around $\chi^0=x^0$ (e.g.\ a Gaussian with a constant, non-zero width $\epsilon$) and a reduced condensate wavefunction $\tilde{\sigma}$ (assumed not to spoil the peaking properties of $\eta$). As a result, the hydrodynamic variables \eqref{eqn:hydrovariables} become functionals of $\tilde{\sigma}$ localized at $\chi^0=x^0$. In particular, one can show that~\cite{Marchetti:2020umh,Marchetti:2020qsq}, when $x^0$ is not too small, $x^0=\langle\hat{\chi}^0\rangle_{\sigma_x^0}\equiv \langle\hat{X}^0\rangle_{\sigma_x^0}/\langle\hat{N}\rangle_{\sigma_x^0}$, where $\hat{\chi}^0$ is the intrinsic scalar field operator associated to the extensive ($1$-body) clock scalar field operator $\hat{X}^0$ given by \eqref{eqn:hydrovariables} with $O(\mathcal{D},\mathcal{D}')=\chi^0$. Thus, the hydrodynamic variables are in general localized with respect to the expectation value of the scalar field clock; as such, they are effective relational observables, see also Section~\ref{sec:relationalism} on the relationalist approach.

One such observable which is of remarkable importance for cosmological applications is the spatial volume, obtained from equation \eqref{eqn:hydrovariables} by choosing matrix elements $O(\mathcal{D},\mathcal{D}')\to V(g_I,g_I')$ via the map between GFTs  and LQG, where $g_I=(g_1,\dots, g_4)\in G^4$~\cite{Oriti:2016qtz}. As suggested by equation \eqref{eqn:hydrovariables}, the behavior of the volume is dictated by the behavior of the (reduced) condensate wavefunction $\tilde{\sigma}$. One can solve the mean-field equations in \eqref{eqn:meanfield} in the mesoscopic regime where GFT interactions are negligible, and show that at late relational times (which dynamically corresponds to large average number of quanta in the condensate, and large volumes), the volume dynamics is equivalent to a flat Friedmann one~\cite{Oriti:2016qtz,Marchetti:2020umh,Jercher:2021bie}; moreover, in this regime, quantum fluctuations on both the volume and the clock variables are typically under control~\cite{Marchetti:2020qsq}. This is truly the classical continuum cosmological regime of the theory. 

At earlier times, instead, the average volume undergoes a quantum bounce for a large range of initial conditions~\cite{Oriti:2016qtz,Marchetti:2020umh,Jercher:2021bie}. This average singularity resolution may however be spoiled by quantum effects both on the volume and on the clock~\cite{Marchetti:2020qsq}. These quantum effects are controlled by the average number of quanta~\cite{Marchetti:2020qsq,Gielen:2019kae,Gielen:2021vdd} and, as one would expect by analogy with condensed matter systems, become important if the latter is small around the average bounce. In this regime, one expects the mean-field approximation (and thus the corresponding hydrodynamic description) to break down.

\paragraph{Recent results and conclusions.}
Given these encouraging results, a substantial theoretical effort has been devoted to (1) the improvement of the approximations used, and (2) the study of more physically realistic systems in order to make contact with observations. Both these directions can be concretely explored quite successfully, as we explain below, which speaks volumes on the solidity and viability of the whole program.
\begin{enumerate}
    \item \textcolor{black}{At the level of the quantum dynamics,} the two most important approximations employed are the mean-field approximation and the approximation of negligible interactions. Though in general going beyond mean-field is far from being easy, a promising strategy might be to follow the Bogoliubov approach to quantum fluids~\cite{pitaevskii2003bose,david}. Including interactions, instead, seems to generally produce cosmic acceleration, which might in turn generate a different, purely quantum geometric mechanism for inflation~\cite{deCesare:2016rsf,Pithis:2016cxg,Jercher:2021bie,tom2}, or a phantom dark energy at very late times~\cite{Oriti:2021rvm}.

    \textcolor{black}{Regarding the implementation of the relational strategy, exact relational observables have been constructed for the first time in the GFT Fock space by incorporating insights gained within the quantum reference frame context (see Section \ref{sec:relationalism}) into the GFT framework~\cite{lucaed}. These relational observables, constructed using frame POVMs, match the effective relational observables described above when the frame measurement associated with the POVMs carries intrinsic uncertainty~\cite{lucaed}.}
    \item There are two broad directions that have been explored in order to make contact with observations. The first is to include more realistic matter fields. As a first step, a scalar field with a non-zero potential has been successfully included~\cite{tom}, leading in particular to interesting insights on the renormalization properties of these models (at least in the emergent, continuum regime).
    The second one consists in moving beyond homogeneity and thus studying the physics of cosmological perturbations~\cite{Gielen:2017eco,Gerhardt:2018byq,Marchetti:2021gcv}. For example, in~\cite{Marchetti:2021gcv} it has been shown that one can match the GR behavior only at late times and for super-horizon perturbations. These results have been substantially improved by causally coupling the reference fields to the quantum geometry and by considering states that include small quantum entanglement between the GFT quanta. The (scalar and isotropic) perturbations emerging from these quantum correlation functions have been shown to satisfy late-times GR dynamics with trans-Planckian quantum corrections~\cite{Jercher:2023nxa,Jercher:2023kfr}. 
\end{enumerate}

\textcolor{black}{In conclusion, from the perspective of the fundamental QG theory, relational cosmological physics effectively emerge as a hydrodynamic phenomenon in the limit in which the average number of quanta in the QG system is large. This section shows that the novel \textit{hydrodynamics on superspace} framework, as advertised by the editors of this collection in the introduction and Section~\ref{sec:benachour}, can be concretely realized for particular TGFT models via a mean-field approximation of the fundamental quantum gravity dynamics (see Section~\ref{sec:thuerigen}) and a relational description of the emergent physics (see Section~\ref{sec:relationalism}). In this way, the work presented in Section~\ref{sec:marchetti} may also serve as a template for extracting cosmological physics from superspace hydrodynamics (including QG effects) from the point of view of other QG approaches.}

\subsection{Emergent Metric Spacetime from Matrix Theory\\\textit{by Robert Brandenberger}}\label{sec:brandenberger}

This contribution starts off with two important messages. Firstly, inflation is not the only early universe scenario  which is consistent with current observational data.  The three criteria for a successful early universe scenario are 1) that the current horizon is much larger than the Hubble radius, 2) that scales which are currently explored in cosmological surveys originate from inside the Hubble radius at some early time, and 3) that there is a mechanism which generated fluctuations with an almost scale-invariant spectrum. These criteria are satisfied in models of rapid acceleration, in certain bouncing cosmologies, and in emergent universe scenarios with thermal fluctuations with holographic scaling~\cite{Brandenberger:2010dk}.

The second main message is that cosmological models with a long period of accelerated expansion cannot be described via effective field theories. Such effective field theories would be non-unitary and would violate the second law of thermodynamics.  In particular, standard inflationary scenarios described at the level of effective field theory are inconsistent~\cite{Bedroya:2019snp,Bedroya:2019tba,Brandenberger:2021pzy}.  As a corollary, we conclude that one needs to go beyond effective field theory is one wishes to have a consistent early universe model.

The second main point of this contribution discusses a proposal to obtain emergent continuous time and space from the BFSS matrix model~\cite{Banks:1996vh}, a proposed non-perturbative definition of superstring theory.  Starting from a high temperature state of this model, a prescription was reviewed for obtaining emergent continuous time and space, see also~\cite{Brahma:2021tkh,Brahma:2022dsd}.  The emergent time extends infinitely into both the past and the future, and the space which emerges is also infinite.  The proposal of~\cite{Brahma:2021tkh,Brahma:2022dsd} yields an emergent spatially flat metric, and the thermal fluctuations yield scale-invariant spectra for both cosmological perturbations and gravitational waves.  This construction can be seen as yielding a realization of the Hagedorn phase of String Gas Cosmology~\cite{Brandenberger:1988aj}.

\section{Discussion and conclusion}\label{sec:conclusion}

The goal of the collection of perspective pieces was to report on interesting new developments in various quantum gravity approaches, theoretical cosmology, quantum foundations and analog gravity, to foster a fruitful exchange of ideas and identify possible common ground for collaborations. 

Next to this, as discussed in the introduction, the superordinate motivation for the editors of this collection was to bring together research directions that, together, form key theoretical pillars as well as potential domains of application of an effective framework to study cosmological physics, incorporating quantum gravity features, given by \textit{hydrodynamics on superspace}. Such hydrodynamics on superspace corresponds to a non-linear and non-local extension of quantum cosmology, it is understood as the result of a coarse-graining of fundamental quantum gravity degrees of freedom, and it is sourced by a growing number of results in mathematical physics and quantum gravity research.

This explained, the collection falls into four thematic units: $(a)$ Map between hydrodynamics and cosmology and analog gravity systems, $(b)$ Phase transitions, continuum limits and emergence in quantum gravity, $(c)$ Relationalism in gravity and quantum gravity and $(d)$ Emergent cosmology from quantum gravity. \textcolor{black}{The content of these units may be seen as standalone perspectives on their respective topics. Beyond this, the results presented therein inform the development of the \textit{hydrodynamics on superspace} framework.} In the following, we briefly summarize key messages from the individual units:

The first thematic unit, Section~\ref{sec:hydrodynamics}, reviewed the main ideas and recent results supporting the existence of a map between hydrodynamics and cosmology on the one hand, and the possibility to realize analog systems for quantum cosmology in the future on the other. The first contribution discussed the shared symmetries between hydrodynamics and symmetry-reduced gravitational superspaces. Crucially, these shared symmetries allow to build a dictionary between the observables of quantum cosmology and the ones of hydrodynamics opening the door for analog models to encode dynamical information. The second contribution summarized the efforts to consider the backreaction of quantum fluctuations on the background in analog gravity. \textcolor{black}{In a sense, this effort goes in the same direction as it aims at encoding effects of the dynamics on the background geometry.} The third contribution finally reviewed the status of analog models tailored to cosmological settings. Importantly, preliminary results on an analog BEC model for reproducing the WDW dynamics show that the geometry of the field space is indeed close to a cosmological spacetime geometry.

The second thematic unit, Section~\ref{sec:phasetransitions}, focused on the continuum limit via coarse-graining and the idea of emergence of spacetime in different quantum gravity approaches. The first contribution critically discussed results on the emergence of a de Sitter-like phase in CDT highlighting evidence that CDT in $3+1$ dimensions in fact lies in the same class of theories as Ho{\v{r}}ava-Lifshitz gravity. The second contribution focused on recent advances regarding coarse-graining and the continuum limit in spin foam gravity via refinement. The third contribution reviewed how this issue is tackled in TGFT using Landau-Ginzburg mean-field theory and emphasized that for realistic models the use of the Lorentz group seems to play a key role to generate a tentative continuum geometry. The last contribution focused on recent results in the asymptotic safety approach and, in particular, discussed how the issue of observables is confronted therein using the relationalist strategy. This thematic unit also shows that to fully characterize different continuum limits in the various approaches it is pressing to construct suitable observables. 

The third thematic unit, Section~\ref{sec:relationalism}, covered recent advancements in developing the relational framework in classical and quantum gravity. The first contribution demonstrated that classically a notion of locality consistent with (i) general covariance, (ii) bulk microcausality, and (iii) the existence of gauge-invariant local subsystems can be preserved in background-independent theories, provided it is defined with respect to dynamical rather than external reference frames. As discussed in the third contribution, such dynamical frames enable a reduced quantization of gravitational systems. It also illustrated how the choice of a dynamical frame influences the physical properties of the resulting quantum gravity model. These effects could be systematically explored using a perspective-neutral framework, where both the frame and the system are quantized, and the perspective of the (quantum) reference frame is adopted only at a later stage. Although implementing this strategy in full quantum gravity remains an open problem, progress has been made in quantum mechanical systems incorporating gravitational elements, as discussed in the second contribution.

The fourth thematic unit, Section~\ref{sec:cosmology}, examined the process of cosmic emergence in quantum gravity, highlighting the challenges inherent in reconstructing continuum cosmological concepts from a pre-geometric quantum gravity framework, as well as the significant impact quantum gravity can have on cosmology by helping to alleviate both conceptual and observational tensions. The first contribution focused on this potential impact, with particular attention to the complex interplay between quantum gravity, inflation, and dark energy. In contrast, the second contribution addressed the challenges associated with emergence itself, specifically the subtle issues surrounding the emergence of time. Finally, concrete examples of cosmic emergence were provided in the third and fourth contributions, within the TGFT approach and superstring theory, respectively. These examples demonstrated how macroscopic properties of the universe can be quantitatively linked to microscopic quantum-gravitational phenomena. \textcolor{black}{In particular, this unit demonstrated that TGFT condensate cosmology is a specific instantiation of the \textit{hydrodynamics on superspace} framework, as advertised by the editors of this collection, and may serve as a template for the extraction of cosmology from superspace hydrodynamics, which encompasses QG signatures, from other quantum gravity approaches.}

To summarize, we hope that this collection represents an important contribution to advancing quantum gravity, cosmology and foundations of physics, and to their overlap, thanks to the lens of \textit{hydrodynamics on superspace} as well as to the many distinct yet interrelated insights offered by the four thematic units. Clearly, this effort transcends traditional boundaries and invites interdisciplinary work, fostering innovation across mathematical physics, condensed matter, quantum gravity and cosmology, with the long-term goal of redefining our understanding of spacetime and the universe.

\subsection*{Acknowledgements}
The work of Jibril Ben Achour is supported by the Sir John Templeton foundation and the MCQST through the seed funding program. Flaminia Giacomini acknowledges support from the Swiss National Science Foundation via the Ambizione Grant PZ00P2-208885, the ETH Zurich Quantum Center, and from the John Templeton Foundation, as part of the ‘The Quantum Information Structure of Spacetime, Second Phase (QISS 2)’ Project. Christophe Goeller acknowledges support by the Alexander von Humboldt Foundation and the MCQST for its support through the seed funding program. Tobias Haas acknowledges support from the European Union under project ShoQC within the ERA-NET Cofund in Quantum Technologies (QuantERA) program and from the F.R.S.- FNRS under project CHEQS within the Excellence of Science (EOS) program. Philipp Höhn acknowledges support of the ID\# 62312 grant from the John Templeton Foundation, as part of the project \href{https://www.templeton.org/grant/the-quantum-information-structure-of-spacetime-qiss-second-phase}{‘The Quantum Information Structure of Spacetime’ (QISS)}. The opinions expressed in this publication are those of the Philipp Höhn and do not necessarily reflect the views of the John Templeton Foundation. Luca Marchetti acknowledges support from the Atlantic Association for Research in Mathematical Sciences. Daniele Oriti acknowledges financial support from the ATRAE program of the Spanish Government, through the grant PR28/23 ATR2023-145735. Daniele Oriti and Andreas Pithis acknowledge funding from the Deutsche Forschungsgemeinschaft (DFG, German Research Foundation) research grants OR432/3-1 and OR432/4-1 and the John-Templeton Foundation via research grant 6242. Antonio D. Pereira acknowledges CNPq under the grant PQ-2 (312211/2022-8), FAPERJ under the “Jovem Cientista do Nosso Estado” program (E26/202.800/2019 and E-26/205.924/2022) and NWO under the VENI Grant (VI.Veni.192.109) for financial support. Andreas Pithis is grateful for the generous financial support by the MCQST via the seed funding through the DFG under Germany’s Excellence Strategy – EXC-2111 – 390814868 and in particular acknowledges funding by the DFG under the author’s project number 527121685 as a Principal Investigator. Sebastian Steinhaus gratefully acknowledges support by the Deutsche Forschungsgemeinschaft (DFG, German Research Foundation) project number 422809950 and Johannes Thürigen is funded by DFG grant number 418838388 and Germany’s Excellence Strategy EXC 2044–390685587, Mathematics Münster: Dynamics–Geometry–Structure.

%

\newpage
\bibliographystyle{JHEP}
\bibliography{references.bib}

\end{document}